\def\comment#1{}
\def\old#1{}
\def\mn#1{*\marginpar[*#1]{*#1}}
\def\mn#1{}
\documentstyle[12pt,epsf]{article}

\newcommand{\sbf}[1]{\mbox{{\scriptsize$\bf{#1}$}}}

 \def\lfrac#1#2{{{{#1}/{#2}}}}
\begin{document}
\begin{center}
\vskip 0.001in
{\Large \bf Crossover from Weak- to
Strong-Coupling Superconductivity
and to Normal State with Pseudogap. }
\vskip 0.4in

{ \bf  E. Babaev}%
\footnote{Emails:
babaev@physik.fu-berlin.de, ~babaev@k-box.ioffe.rssi.ru; ~~~~
 kleinert@physik.fu-berlin.de ~~ URL:
http://www.physik.fu-berlin.de/\~{}kleinert ~~ Phone/Fax:
 0049 30 8383034.
}${}^{,2}$
  and { \bf H. Kleinert${}^1$ }

\vskip0.5cm

  (1) \ Institut f\"ur Theoretische Physik \\
Freie Universit\"at Berlin, Arnimallee 14, 1000 Berlin 33, Germany

  (2)  \ A.F.Ioffe Physico-Technical Institute \\
 Russian Academy of Sciences,
Politechnicheskaja str. 26,
St. Petersburg, 194021, Russia

%{\em    }
\vskip 0.15in

\end{center}
\begin{abstract}
For
 an electron gas with
a $ \delta $-function attraction we investigate the
crossover from weak-coupling to
strong-coupling superconductivity
as well as normal state near the temperature $T^*$,
at which
the strong coupling produces
a pseudogap
in the energy spectrum due to the binding of electron pairs.
We present curves
for the behavior of the
superconductive
transition temperature, the gap formation temperature,
 the gap size and several thermodynamic quantities
 as  functions of coupling strength
and temperature, both in two and three dimensions.
%The results should be useful
%for interpreting experimental data in
%underdoped and optimally doped cuprates.
\old{In three - dimensional case
we propose proliferation transition of three dimensional XY-model
for describing superconductive transition in strong-coupling limit
of the theory.
The relevant thermodynamic
functions are calculated for future use, in particular
for a possible comparison
with experimental results in
 underdoped and optimally doped cuprates.
}
\end{abstract}
\baselineskip 0.75cm

\section{Introduction}

The crossover from BCS superconductors  to a
 Bose-Einstein condensate of tightly bound
fermion pairs was first  studied
many years ago in
Refs.~\cite{Ea}--\cite{Le}.
Sparked by experimental studies  of
short coherence length   cuprate
superconductors,
it has recently  attracted  renewed interest
\cite{Uem}- \cite{ek},
 the experiments
show  an anomalous behavior in the normal phase well
{\em above\/}  the superconductive transition
$T_c$ (see the review \cite{Pi}, also
\cite{ranrev}, \cite{Levi}-\cite{ang3}).
Anomalous is
 the temperature dependence
of  resistivity, specific heat, spin susceptibility, etc.\,.
Moreover, angular-resolved photoemission spectroscopy experiments  (ARPES )
indicate the existence of a {\em pseudogap\/}
in the single-particle excitation spectrum
 \cite{ang1} - \cite{ang4}
manifesting itself in a significant   suppression
of low-frequency spectral weight
 above $T_c$,
 this being similar to the
complete suppression below $T_c$ in an ordinary superconductor
due to the energy gap.

In essence, such anomalous  properties of the normal state
of superconductor
can be described
by a simple model of
superconductivity in which electrons are bound to pairs
by a $ \delta $-function potential.
We merely must leave the BCS regime and
go to stronger couplings \footnote{
It should be noted that phase diagram
in the case of such complicated materials as
High-$T_c$ cuprate superconductors
contains large region where
anomalous properties
of normal state are
governed by
antiferromagnetic correlations.
}.

In this paper we present a detailed study of
the crossover in such a model.
Physically, the most important distinctions
between conventional (BCS) and strong-coupling (Bose-Einstein)
regime
lies in the fact that in the former
only a small fraction of the
conduction electrons
is paired
with
the superfluid density involving all pairs, whereas
in the latter
practically all  carriers are paired
below a certain temperature $T^*$, although not condensed,
effect that results in deviation from the Fermi liquid behavior
in the region between $T^*$ and $T_c$.
The temperature
has to be lowered further below some critical temperature
$T_c < T^*$ to make
these pairs condense and
establish
 phase coherence, which leads to
superconductive behavior.
We shall neglect the coupling to the
magnetic vector potential throughout the upcoming
discussion, so that
the phase coherence below
$T_c$ can be of long range, unspoiled by the Meissner effect
which would reduce the range to
a finite
penetration depth.

The pseudogap behavior between
 $T_c$
and  $T^*$ is characterized  by
short-range pair correlation functions.
The common physical origin of
the
superconductive gap below $T_c$
and the pseudogap
above $T_c$ observed in cuprates
is suggested by
the
above-quoted ARPES
data, which
show that
the two gaps
have the same
 magnitude and wave vector dependence.
Important experiments on the gap properties are:
\comment{
Further experimentally observed
  properties of the pseudogap of pseudogap state
above $T_c$ can be in principle
be features of  precursor pairing models
even though  there are no reliable experimental
results yet about the nature
of the physics by which the pseudogap state
in cuprates is governed}
\old{
\footnote
{The authors of \cite{ek} suggest that
in cuprates there are two characteristic
crossovers above $T_c$, the lowest one
slowly decreasing with decreasing carrier
density, which in the dilute limit seems to be related to the transition
studied in our paper,
being a manifestation of the separation
of the pair formation temperature and
temperature of the onset of phase coherence.
The other crossover temperature
grows
quickly with decreasing doping
in the dilute regime and is apparently
due
to the fact that high-$T_c$
superconductor
materials are
doped antiferromagnetic insulators.}}
\begin{itemize}
\item[1.]
In experiments on YBCO \cite{cond1},
\cite{cond2}, a significant suppression of
in-plane conductivity $\sigma_{ab}(\omega)$
 was observed at frequencies below 500 ${\rm cm}^{-1}$  beginning
at temperatures much above $T_c$.
Experiments \cite{dc1}, \cite{dc2} on underdoped
samples revealed
deviations from the linear resistivity law. In particular,
$\sigma_{ab}(\omega=0;T)$
increases slightly with decreasing $T$
below a certain temperature.
\item[2.]
Specific heat experiments \cite{sp} also clearly display pseudogap behavior
much above $T_c$.
\item[3.]
NMR and neutrons observations in \cite{u5} and \cite{u6}
show that below temperatures $T^*$ much higher than $T_c$,
 spin susceptibility starts decreasing.
\old{Within the model  to be studied the
connection of  pseudogap and loss of magnetic response
was studied theoretically in \cite{shar}.}
\item[4.]
Experiments on  optical conductivity
\cite{opt1}, \cite{opt2} and tunneling exhibit the opening of a pseudogap.
A  review of actual experimental data confirming the pseudogap behavior
of the underdoped and optimally doped cuprates
is given in \cite{ranrev}, \cite{opt1}.
\end{itemize}

\old{
Experimental evidence
for
of the phase separation in modern
high-$T_c$
superconductors
is best seen on a schematic plot
in Fig.~\ref{exp},
taken from the experimental
work in Ref.~\cite{opt1}.%
\begin{figure}[htpb]
%\leavevmode
\vspace{.3cm}
\epsfxsize=0.5\columnwidth
\centerline{\epsffile{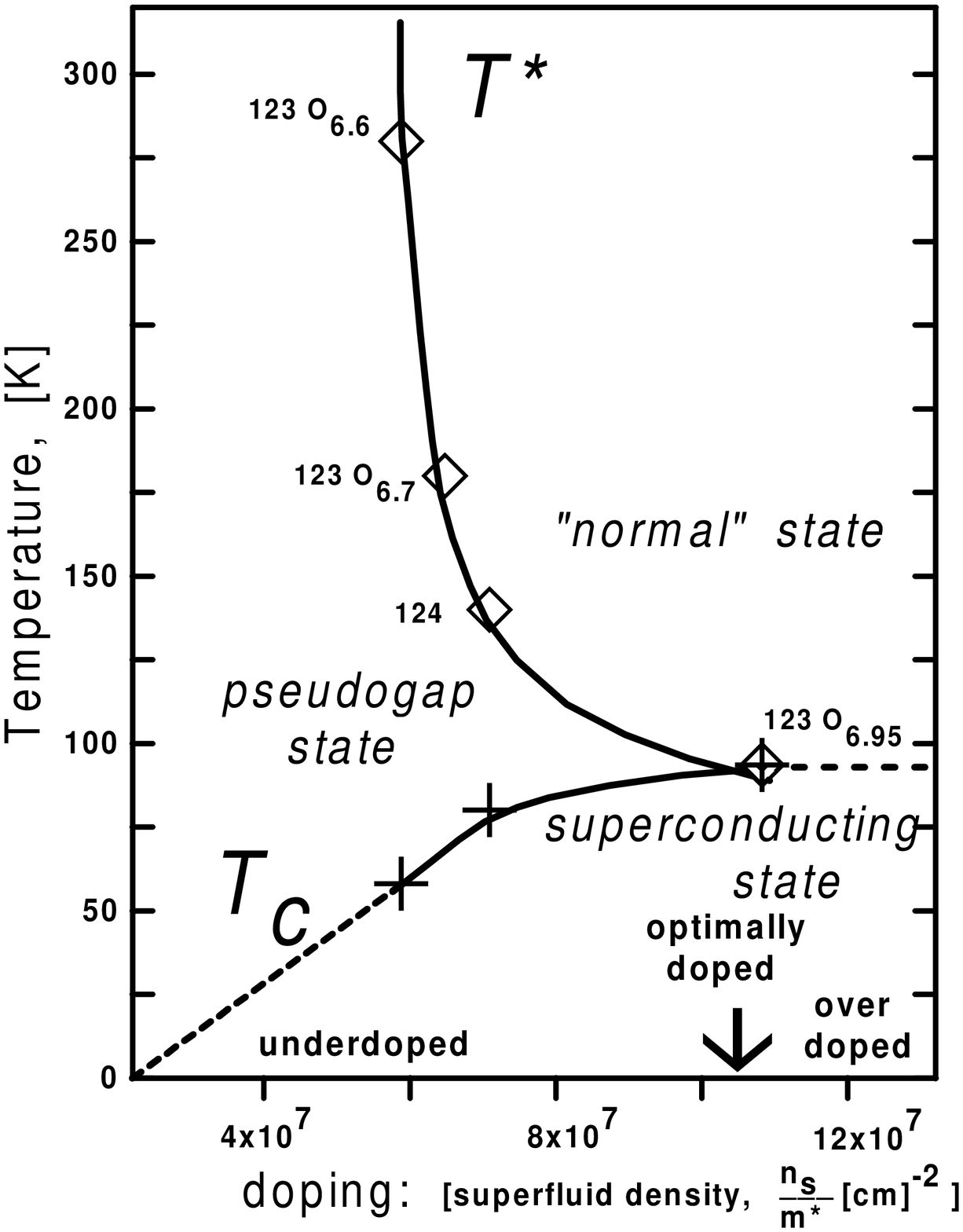}}
\caption{Schematic phase diagram of the cuprate superconductors
taken from Ref.~[35]. In the underdoped regime,
 a pseudogap state forms between the temperatures
$T_c$ and $T^*$. The two curves
merge at an optimal doping
where the pseudogap and the superconducting gap
form
at the same
temperature. The upper temperature $T^*$ is
determined
by measuring the c-axis
conductivity
and, while the
doping level follows from
measurements of the superfluid density
%$\omega_{\ps ps}^2 = $
$n_s/m^*$ in the CuO$_2$ planes.
}
\label{exp}
\end{figure}}

In the model to be investigated in this paper,
the
crossover
from BCS-type to Bose-type
superconductivity and pseudogap state
will take place
either by
varying the coupling
strength, or by
decreasing
the carrier density.

Analytic calculations
will be performed
using a crossover parameter $x_0$. This parameter
is directly related to the ratio of chemical potential and
gap function at zero temperature. It
is a monotonous function of coupling strength and
carrier density
[to be seen in Figs.~\ref{f1}
and \ref{f2} in two and three dimensions, respectively].
It is also a
direct measure for the
scattering length $a_s$ of the electron-electron
interaction, at zero temperature
 in three dimensions [to be seen in Eq.~(\ref{1.8})], and
to the binding energy of the
electron-electron pairs in two dimensions [to be seen in
Eq.~(\ref{@binding})].
\old{ound state energy
of constant while coupling varies
it is easy to see from Eq. (\ref{1.8}), (\ref{q1})
and corresponding figures (\ref{f3a}), (\ref{f3b}) that
our crossover parameter $x_0$ can be changed
as well simply by variation of the density of carriers
$n(T=0)$ and "weak-" and "strong-"
coupling limits can be then reached
in the dilute- and high-density-limits
correspondingly.                               }

\old{
The simple model to be studied here
is not really capable of
representing the complete physics
in a high-temperature superconductor.
It is, however, hoped that the model
displays
some essential features
of the observed phenomena.
It therefore seems
useful to study this model as well as possible.}

Our analysis of the model start with
a mean-field approximation
to the collective pair field theory \cite{6'}.
%This approximation by itself is not reliable
%in the strong coupling regime near the temperature
%$T^*$ where the pairs and pseudogap form, due to strong fluctuations.
%In particular, the smoothness of the
%behavior near $T^*$ is not matched by the
%second-order transition obtained in the
%mean-field approximation
%of our model.
%It is, however, believed that the mean-field {\em value\/} of
%$T^*$ will remain
%nearly unchanged by these fluctuations.
It is well known,
that mean-field results are reliable
at all temperatures
for weak coupling strength, i.e. in the BCS regime.
As this regime
is approached from the strong-coupling side,
the temperature $T^*$ where pairs are formed and the
temperature $T_c$ where phase coherence
sets in
merge to the single BCS phase transition temperature $T_c$.
This merging
will be described analytically in this paper in two as well as three
dimensions.

Mean-field results are also reliable
at strong couplings
if the
temperatures are sufficiently small to suppress fluctuations [see the
discussion
in \cite{RR8}].

Apart from that,
mean-field results at stronger couplings seem to indicate
correctly
the {\em position\/} of the temperature $T^*$ where pseudogap forms due to
precursor
  pair formation
\cite{R8}.
The model shows the binding of noncodencerd pairs
by the appearance a nonzero
complex gap function ${\bf
\Delta}(T)$.
The precise temperature behavior near $T^*$
is certainly predicted wrongly
as being
$\propto (T^*-T)^{1/2}$. This would suggests a second-order
phase transition, whereas the experimental
data show a smooth crossover phenomenon.
Here the effect of fluctuations
is too important to be calculable analytically.

In this paper we shall focus attention
upon
the onset of long-range order when lowering the temperature down from $T^*$.
Due to the strength of the coupling,
this regime lies outside
the mean-field approximation.
It appears, however, that
by extracting the lowest gradient terms
governing the Gaussian fluctuations
around the mean-field
we shall
obtain sufficient information
to study the onset of long-range order
and its destruction at $T_c<T^*$.

In three dimensions, this
was first investigated  in Ref.~\cite{Noz}
by summing particle-particle ladder diagrams
which
correspond to
Gaussian fluctuations
around the mean field, in the functional integral formalism
it was studied in Ref.~\cite{R8}.

As it was noted above, at strong couplings,  the formation of a gap does not
imply
the existence of superconductive currents.
Although the modulus of the gap field
$  \Delta({\bf x})\equiv |{\bf \Delta}({\bf x})|$ is nonzero,
the phase of $ {\bf \Delta}$ will initially fluctuate
so violently that
 long-range order cannot be established.
This phenomenon is well-known from
XY-models [O(2) classical Heisenberg model],
which describe the fluctuations of a field of two-dimensional
unit vectors ${\bf n}=(\cos \phi,\sin \phi)$.
The  fluctuations of the phase angle $\phi$
prevent the existence of a long-range order
above a certain temperature.
%In  two dimensions, this temperature
%is $T_{KT}$, the Kosterlitz-Thouless temperature.
%whose magnitude can be understood from
%the Coulomb gas properties exhibited by this model in
%its dual description \cite{GFCM}.
%In three dimensions, there exists an equivalent description
%in terms of a gas of vortex lines \cite{GFCM}.

%With the help of the
%calculated gradient terms
%we shall be able
%to set up two- and three-dimensional  XY-models
%for the fluctuations in the superconductor at strong couplings
%and find the onset of phase coherence in two and three dimensions.
%The plan of the paper is following:
% The different properties of size and phase fluctuations
%was also exploited in Refs.~\cite{Po} and \cite{Trav}.
Much of the previous work on the model with an attractive $\delta$-potential
at strong couplings
was devoted
to zero-temperature  or
to a study of the critical region
near $T_c$
where phase coherence sets in.
%The weak- to strong-coupling crossover
% of the Kosterlitz-Thouless in two dimensions
% was studied
%in Refs.~\cite{Dr}, \cite{shar}.
Apparently,
there is no analytic work
displaying a complete
set of global properties
of the
pseudogap state
above $T_c$.
There only exist some
numeric results
in Ref.~\cite{shar}
on the
paramagnetic
susceptibility
in the pseudogap phase
of two-dimensional superconductors,
derived within
the same
model after adding a paramagnetic term
to Hamiltonian \footnote{ For  Monte-Carlo study
of the normal state just above $T_c$ see Ref. \cite{rQMC}}.
In the present paper, we shall derive as well
analytically
the  behavior  for all coupling strengths
of the pseudogap as well as
the thermodynamics of the pseudogapped phase in both two and three dimensions.

The weak- to strong-coupling crossover
 of the Kosterlitz-Thouless in two dimensions
 was studied
in Refs.~\cite{Dr}, \cite{shar}.
In Ref.~\cite{shar}
it was investigated within
the same
model as ours at a fixed carrier density,
 but only
numerically\footnote{We shall see in Section~4
that these numerical results do not cover
the entire
crossover region, in particular, the merging of $T_{KT}$
and $T^*$ in the weak-coupling region  is missing - the
effect that we show analyticaly in our paper.}
{}.
The different properties of size and phase fluctuations
was also exploited in Refs.~\cite{Po} and \cite{Trav}.

%Our results derived in the weak-coupling limit of the theory
%are expected to be reliable
%for  weak and moderately strong
%couplings slightly beyond the BCS regime.

The plan of the paper is the following:

We begin by reproducing
the  results
of Refs.~\cite{M},
 \cite{2Danalit} for the gap $ \Delta (0)$
and chemical potential at zero temperature.
These are subsequently extended
by
equations for the
temperature behavior of  gap and pseudogap, as well
as of thermodynamic functions.

In   Section~2  we calculate the crossover
from  BCS to strong-coupling superconductivity
at small  but finite temperatures and present solutions
for the gap and thermodynamic functions of the superconductive state.

In  Section~3 we study  the crossover
from   BCS superconductivity near $T_c$
to the onset of pseudogap behavior
near $T^*$.

In  Sections ~4 and ~5
we go beyond mean-field approximation
and study crossover of superconductive transition
in both two and three dimensions.
In
the strong-coupling limit of
superconductors,
we set up an equivalent $XY$-model
in two as well as three dimensions
with the help of a gradient expansion for the phase of the order parameter.
This allows us to find the onset of phase coherence
in a non-perturbative way.
{ \rm Even though
in two dimensions, this was done before
via
 Kosterlitz-Thouless arguments we  argue that
merging of temperatures of pair formation and
condensation was missed.
 In three dimensions, our discussion is new.
Previous results were derived form a study
of the condensation of the gas of composite bosons
via retaining corrections to the number equation.

In the  weak-coupling  limit of both two- and three-dimensional
supercondictors
we show how the
$XY$-model transition
converges with the transition
in the
BCS theory.

\section[Crossover from BCS to Strong-Coupling Superconductivity  Near Zero
Temperature]
{Crossover from BCS to Strong-Coupling \\Superconductivity  Near Zero
Temperature}

The Hamiltonian of our model is the typical BCS Hamiltonian
in $D$ dimensions ($\hbar=1$)
\begin{eqnarray}
H &=& \sum_\sigma \int \! d^D x
        \, \psi_\sigma^{\dag} ({\bf x})
        \left(-{{\bf \nabla}^2 \over 2m} -\mu\right)
        \psi_\sigma({\bf x})
         + g \int\!d^D x\,
        \psi_\uparrow^{\dag}({\bf x}) \psi_\downarrow^{\dag}({\bf x})
        \psi_\downarrow^{\phantom{\dag}}({\bf x})
        \psi_\uparrow^{\phantom{\dag}}({\bf x})
        \label{1.0},
\end{eqnarray}
where $\psi_\sigma({\bf x})$ is the Fermi field operator,
$\sigma=\uparrow,\downarrow$
denotes the spin components,
$m$ is the effective fermionic mass, and
$g < 0 $  the strength of an attractive potential
$ g  \delta ({\bf x} - {\bf x}')$.

The mean-field equations for the
gap parameter $\Delta$ and the chemical potential $\mu$
are obtained  in the standard way from the equations
 (see for example \cite{6'}, \cite{shar}
and Appendix~A):

\begin{eqnarray}
-{1\over g} &=& \frac{1}{V} \sum_{\bf k} {1\over 2 E_{\bf k}}
\tanh{E_{\bf k} \over 2T} ,\label{1.1}\\
  n &=& {1\over V} \sum_{\bf k} \left(1-{\xi_{\bf k}
 \over E_{\bf k}} \tanh{E_{\bf k} \over 2T}\right),
\label{1.2}
\end{eqnarray}
where the sum runs over all
wave vectors
$\bf k$,
 $N$ is the total number
of fermions,
 $V $  the
volume of the system,
 and
\begin{equation}
 E_{\bf k}=\sqrt{\xi_{\bf k}^2 + \Delta^2}
{}~~~\mbox{with}~~~
\xi_{\bf k} = {{\bf k}^2 \over 2 m} - \mu
\label{1.3}
\end{equation}
 are the energies of single-particle excitations.

Changing the  sum over {\bf k} to  an integral
over $\xi$
and over the directions of ${\bf k}$, on which the integrand does not depend,
 we arrive
in three dimensions at the gap equation:
\begin{equation}
{1 \over g} = \kappa_{3} \int_{-\mu}^\infty d\xi {\sqrt{\xi+\mu}
\over 2 \sqrt{\xi^2+\Delta^2}} \tanh{\sqrt{\xi^2 +\Delta^2} \over 2 T}
\label{1.3.1},
\end{equation}
where the constant $\kappa_{3}= m^{3/2} / \sqrt{2} \pi^2$ has dimension
energy$^{-3/2}$/volume.
In two-dimensions, the  density of
states is constant, and the gap equation becomes
\begin{equation}
{1 \over g} = \kappa_{2} \int_{-\mu}^\infty d\xi  { 1
\over 2 \sqrt{\xi^2+\Delta^2}} \tanh{\sqrt{\xi^2 +\Delta^2} \over 2 T}
\label{1.3.2},
\end{equation}
with a constant $ \kappa_{2}= m/2\pi$  of dimension energy$^{-1}$/two-volume.
In two dimensions, the particle number in Eq.~(\ref{1.2}) can
be integrated
with the result:
\begin{equation}
n=\frac{m}{2 \pi} \left\{
\sqrt{\mu^{2} + \Delta^{2}} + \mu +
2 T \log{\left[
1 + \exp{\left(-\frac{\sqrt{\mu^{2} + \Delta^{2}}}{T}\right)}
\right]}\right\},
\label{numb}
\end{equation}
the right-hand side being a function $n(\mu, T, \Delta) $.

The $\delta$-function potential produces an
artificial divergence and requires
 regularization. A BCS superconductor possesses
 a natural cutoff supplied by the Debye frequency $ \omega _D$.
For the crossover problem
to be treated here
this is no longer a useful quantity, since in the strong-coupling
limit all
 fermions
participate in the interaction, not  only those
in a thin shell of width $ \omega _D$ around the Fermi surface.
To be applicable in this regime,
 we  renormalize
the gap equation in three dimensions with the help of the
experimentally observable
$s$-wave scattering length $a_s$,
for which the  low-energy limit of the
two-body scattering process gives an equally divergent
expression \cite{h}--\cite{r2}:
\begin{equation}
        {m \over 4 \pi a_s}
=
{1\over g}
        + {1\over V}
\sum_{\bf k}
{m \over {\bf k}^2} .
 \label{1.4}
\end{equation}
Eliminating $g$ from (\ref{1.4}) and  (\ref{1.1})
we obtain a renormalized gap equation
\begin{equation}
        -{m \over 4 \pi a_s} = {1\over V} \sum_{\bf k}
        \left[{1\over 2 E_{\bf k}} \tanh{E_{\bf k} \over 2T}
 - {m \over {\bf k}^2} \right],
        \label{1.5}
\end{equation}
 in which $1/k_Fa_s$ plays the role
of a dimensionless coupling constant which monotonically increases
 from $-\infty$ to $\infty$ as the bare
coupling constant $g$ runs from small
(BCS limit) to  large values
(BE limit).
This equation is to be solved
simultaneously with (\ref{1.2}).
These mean-field equations were
first analyzed
at a fixed carrier density in Refs.~\cite{R8} and \cite{R-rev}.
Here we shall first
reproduce the obtained
estimates for $T^*$ and $\mu$.

In the BCS limit, the chemical potential $\mu $
does not differ much
from the Fermi energy
 $\epsilon_F$, whereas with increasing interaction strength,
  the distribution function $n_{\bf k}$
 broadens and $\mu $ decreases, and
in the BE limit, on the other hand  we have tightly bound
pairs and nondegenerate fermions with a large negative chemical
potential $|\mu|\gg T$. Analyzing  Eqns.~  (\ref{1.2}) and (\ref{1.5})
we have from (\ref{1.2}) for the critical
temperature in the BCS  limit ($\mu \gg T_c$)
$T_c^{\rm BCS}=8e^{-2}e^\gamma\pi^{-1}\epsilon_F \exp(-\pi/2k_F|a_s|)$
where $\gamma=- \Gamma '(1)/ \Gamma (1) = 0.577 \dots~$,
from (\ref{1.5}) we have that chemical
potential in this case is $\mu = \epsilon_F$.
In the strong Eq.~ (\ref{1.5}) determines $T^*$,
whereas Eq.~ (\ref{1.2}) determines $\mu$. From
Eq~. (\ref{1.2}) we obtain that in the BE limit $\mu = - E_b/2$,
 where $E_b=1/m a_s^2$ is the binding energy of the bound pairs.
In the BE limit, the pseudogap sets in at
$T^* \simeq E_b/2 \log(E_b / \epsilon_F)^{3/2}$.
A simple ``chemical'' equilibrium estimate
$(\mu_b=2\mu_f)$ yields  for the temperature
of pair
 dissociation: $T_{\rm dissoc} \simeq E_b/\log(E_b/\epsilon_F)^{3/2}$
which shows at strong couplings
$T^*$ is indeed related to  pair formation   \cite{R8},
\cite{RR8} (which in the strong-coupling regime
lies above  the temperature of
phase coherence  \cite{Noz}-\cite{ranrev}).

The gap in the spectrum
of  single-particle excitations has a special feature \cite{Le}, \cite{Ea},
\cite{R-rev}
when the chemical potential changes its sign.
The sign change occurs at the minimum of the Bogoliubov
quasiparticle energy $E_{\bf k}$ where this energy
defines the gap energy in the quasiparticle spectrum:
\begin{eqnarray}
E_{\rm gap}={\rm min}\left(\xi_{\bf k}^2 +\Delta^2 \right)^{1/2} .
\label{1.5.1}
\end{eqnarray}
Thus, for positive chemical potential,
the gap energy is given directly by the gap function $ \Delta$, whereas for
negative chemical potential,
it is larger than that:
\begin{eqnarray}
E_{\rm gap}=  \Biggl\{ \matrix{
\Delta & {\rm for}  \ \ \ \mu >0 ,\cr
(\mu^2 +\Delta^2)^{1/2} & {\rm for} \ \ \ \mu < 0 .\cr}
\label{1.5.2}
\end{eqnarray}
In three dimensions at $T=0$, equations (\ref{1.5}), (\ref{1.2})
were solved analytically
in entire crossover region in \cite{M} to obtain
$\Delta$ and $\mu$ as functions of
crossover parameter $1/k_Fa_s$. The results are
\begin{eqnarray}
        {\Delta \over \epsilon_F}
        &=&
        {1\over [x_0 I_1(x_0) + I_2(x_0)]^{2/3}}   ,
        \label{1.6}
        \\
        {\mu \over \epsilon_F}
        &=&
        {\mu \over \Delta} {\Delta \over \epsilon_F}
        = {x_0 \over [x_0 I_1(x_0) + I_2(x_0)]^{2/3}},
        \label{1.7}
        \\
        {1\over k_F a_s}
        &=&
        - {4\over \pi} {x_0 I_2(x_0) - I_1(x_0)
        \over [x_0 I_1(x_0) + I_2(x_0)]^{1/3}}        ,
        \label{1.8}
\end{eqnarray}
with the functions
\begin{eqnarray}
%       \nonumber
        I_1(x_0)
        &=&
        \int_0^\infty \!\!\!\! dx
        {x^2 \over (x^4 - 2 x_0 x^2 + x_0^2 +1)^{3/2} }
        \nonumber \\
        &=&
        (1+x_0^2)^{1/4}
        E({\scriptstyle \pi\over \scriptstyle 2},\kappa) - {1\over 4
x_1^2(1+x_0^2)^{1/4}}
        F({\scriptstyle \pi\over \scriptstyle 2}, \kappa), \nonumber \\
        \label{1.9}\\
        I_2(x_0)
%        &=&
%        -{1\over2} \int_0^\infty\!\!\!\! dx\, x
%        {d \hphantom{x}\over dx}
%        {1\over E_x} = {1\over 2} \int_0^\infty\!\!\!\!dx {1\over E_x}
%        \nonumber \\
        &=&
        {1\over 2} \int_0^\infty\!\!\!\! \,dx
        {1\over (x^4 -2x_0 x^2 + x_0^2 +1)^{1/2}}
        \nonumber \\
        &=&
        {1\over 2 (1+x_0^2)^{1/4}}
        F({\scriptstyle\pi\over \scriptstyle 2}, \kappa)  ,
        \label{1.10}\\
        \kappa^2
        &=&
        {x_1^2 \over (1+x_0^2)^{1/2}}                      ,
        \label{1.11}\\
        x^2
        &=&
       {k^2 \over 2m}{1 \over \Delta} \ , \ \  \ \ x_0= {\mu \over \Delta},
   \ \ \ \ x_1 = \frac{\sqrt{1+x_0^2}+x_0}{2} ,
        \label{1.12}
\end{eqnarray}
and $E({\pi\over 2},\kappa)$ and $F({\pi\over 2},\kappa)$ are
the usual elliptic integrals. The quantities
(\ref{1.6}) and (\ref{1.7}) are plotted
as functions of the crossover parameter $x_0$
in Fig.~\ref{f1}.
\begin{figure}[tbh]%##
\input 3d0.tps
\caption{
Gap function $\Delta$
and chemical
potential $\mu$ at zero temperature
as functions of $x_0$ in three dimensions.
}
\label{f1}\end{figure}

In two dimensions, a nonzero bound state energy $\epsilon_0$
exists for any coupling strength. The cutoff
can therefore be eliminated by
subtracting from the two-dimensional zero-temperature gap equation
\begin{equation}
- \frac{1}{g} = \frac{1}{2V} \sum_{\bf k} \frac{1}{\sqrt{\xi_{\bf k}^2 +
\Delta^2}}=
\frac{m}{4 \pi} \int_{- x_0}^{\infty} d z \frac{1}{\sqrt{1+ z^2}},
\end{equation}
where $ z=k^2/2m\Delta-x_0$,
the bound-state equation
\begin{equation}
-\frac{1}{g}=\frac{1}{V}\sum_{\bf k} \frac{1}{ {\bf k}^2/m + \epsilon_0}=
\frac{m}{2\pi}\int_{-x_0}^\infty d z \frac{1}{ 2z+\epsilon_0/\Delta+2x_0}.
\label{@binding}\end{equation}
After performing the elementary integrals,
we find:
\begin{equation}
\frac{\epsilon_0}{\Delta}=\sqrt{1+x_0^2} - x_0.
\label{@gal2D}\end{equation}
{}From Eq.~(\ref{numb}) we see that at zero temperature,
gap and chemical potential
are related to $x_0$ by
\begin{eqnarray}
   {\Delta \over \epsilon_F }
   &=&
   {2 \over x_0 +\sqrt{1+x_0^2}},
   \label{1.13}
\end{eqnarray}
\begin{eqnarray}
   {\mu \over \epsilon_F}
   &=&
   {2 x_0 \over x_0 +\sqrt{1+x_0^2}}.
   \label{1.14}
\end{eqnarray}
The two relations are plotted in Fig.~\ref{f2}.
Combining (\ref{@gal2D}) with (\ref{1.13})
we find
the dependence of the ratio $ \lfrac{\epsilon_0}{\epsilon_F} $
on the crossover parameter $x_0$:
\begin{equation}
\frac{\epsilon_0}{\epsilon_F}=
2 \frac{\sqrt{1+x_0^2} -x_0}{\sqrt{1+x_0^2} +x_0 }
\label{q1}
\end{equation}

\begin{figure}[tbh]
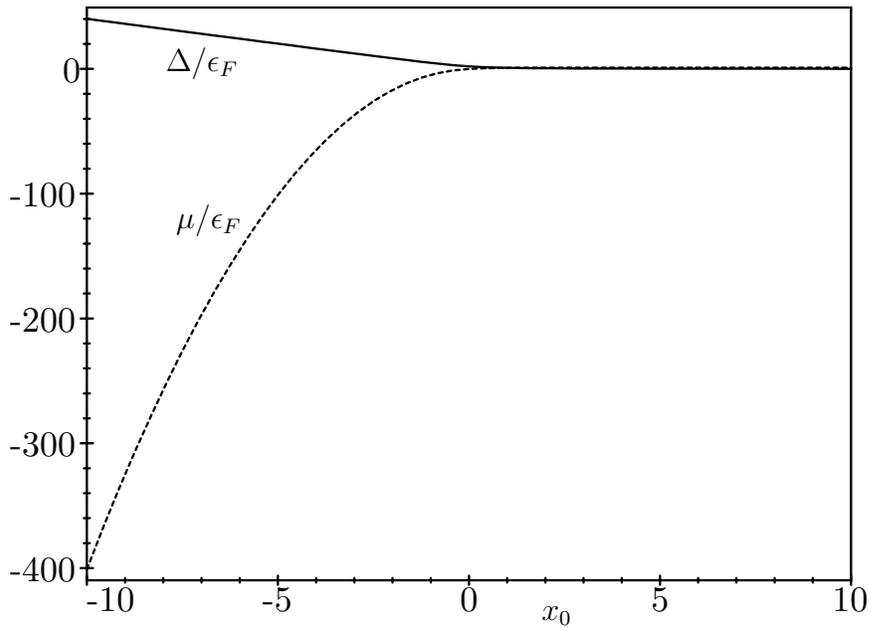
%##
\input 2d0.tps
\caption{
Gap function $\Delta$
and chemical
potential $\mu$ at zero temperature
as functions of $x_0$
in two dimensions.
}
\label{f2}\end{figure}

We have extended
all these relations to non-zero temperature.
For this purpose, we
do not fix the carrier density
but assume
the presence of a reservoir which provides us with a
temperature-independent chemical potential
$\mu=\mu( 1/k_Fa_s ; T=0 )$.
Such a fixed $\mu$
will be most convenient
for deriving simple analytic
results for the finite-temperature behavior of the system.%
\old{
Actually, experiments do
not exhibit
a
temperature-independent carrier density either,
since
many
low-carrier
density  high-temperature superconductors
are
doped insulators with
a   relatively high
crossover temperature $T^*$ to the anomalous normal phase.
When doping is very weak,
these
possess a certain amount of
temperature dependence of the
 carrier density due to ionization of impurities}%
\footnote{In Ref.~\cite{shar},
the temperature dependence of the
chemical potential
was calculated numerically
within a "fixed carrier density model", where it turned out to be very small
in comparison with the dependence on the coupling strength.}

In our calculation we use $x_0$ as the most convenient
 crossover parameter,
since it depends
via
the simple relation (\ref{1.12})
on the chemical potential which
can be measured  rather directly experimentally \cite{Rie}.
The parameter $x_0$
ranges from $-\infty$ in the strong-coupling (Bose-Einstein) limit
to $\infty$ in the weak-coupling (BCS) limit.
The relation between  $x_0$
 and the inverse reduced coupling strength
between the electrons $1/k_F a_s$
is plotted
for three-dimensional system
in Fig.~\ref{f3a}.
The corresponding relation
(\ref{q1}) in two dimensions
between $x_0$ and
the bound state energy $\epsilon_0$
of the electron pairs
is plotted on Fig.~\ref{f3b}.

\begin{figure}[tbh]
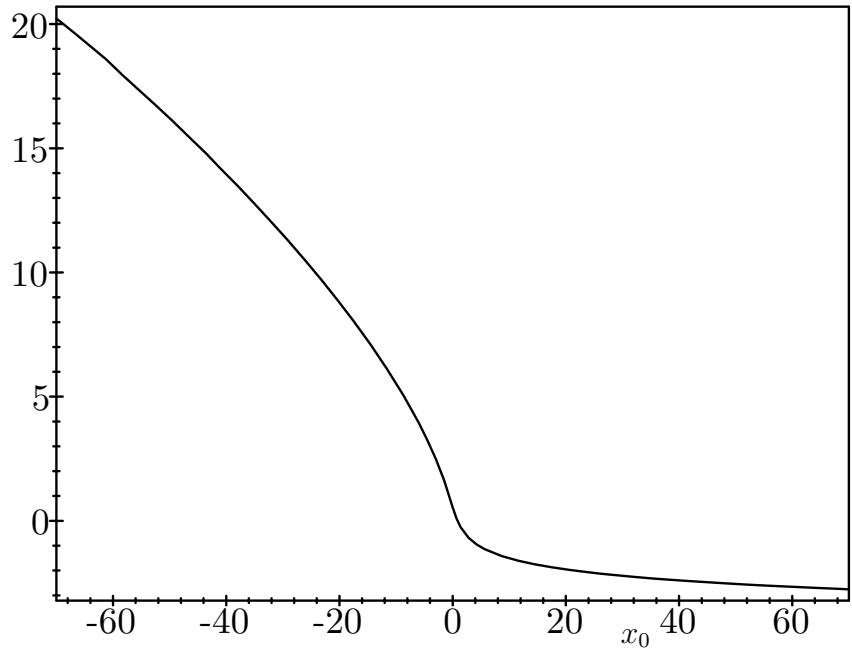
%##
\input as.tps  \\[-2cm]
\caption{
Dependence of $1/k_F a_s$ on the
crossover parameter $x_0$ in two dimensions.
}
\label{f3a}\end{figure}

\begin{figure}[tbh]
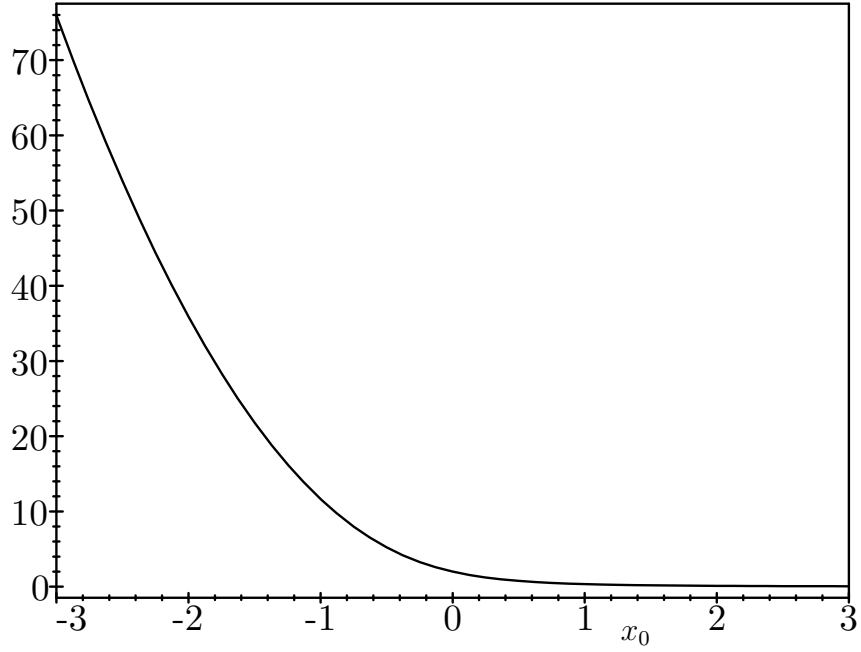
%##
\input e0.tps  \\[-2cm]
\caption{
Dependence of $\epsilon_0/\epsilon_F$ on the
crossover parameter $x_0$ .
}
\label{f3b}\end{figure}

In Fig.~\ref{f3} shows
the temperature behavior of $\Delta$ near $T=0$ for
different coupling strengths in three dimensions.
In
Fig.~\ref{f4} does the same thing
in
 two dimensions.
Figures~\ref{f5} and \ref{f6} display
dependence
of the temperature $T^*$ where the gap vanishes
on the
coupling strength
parameter $x_0$.

As was noted above, recent experiments
on the underdoped cuprates showed that an
ordinary superconductive gap
develops smoothly to a pseudogap
above $T_c$. Within the mean-field approximation
we obtain the analytic results shown in Fig.~\ref{f3}
and \ref{f4}.

\begin{figure}[tbh]%##
\input 3dt.tps   \\[-2cm]
\caption{
Temperature dependence of gap function
in three dimensions.
Solid line corresponds to crossover parameter $x_0=10$ (i.e., in the BCS
regime),
the crosses to $x_0=0$ (i.e., in the intermediate regime),
lines with boxes and circles represent $x_0=-2$ and $x_0=-5$ cases
correspondingly
and the dashed line corresponds to $x_0=-10$
(i.e., in strong-coupling regime).
}
\label{f3}\end{figure}

\begin{figure}[tbh]%##
\input 2dt.tps  \\[-2cm]
\caption{
Temperature dependence of gap function
in
two dimensions.
Solid line corresponds to crossover parameter $x_0=10$ (i.e., in the BCS
regime),
the crosses to $x_0=0$ (i.e., in the intermediate regime),
lines with boxes and circles represent $x_0=-2$ and $x_0=-5$ cases
correspondingly
and the dashed line corresponds to $x_0=-10$
(i.e., in strong-coupling regime).
}
\label{f4}\end{figure}

\begin{figure}[tbh]
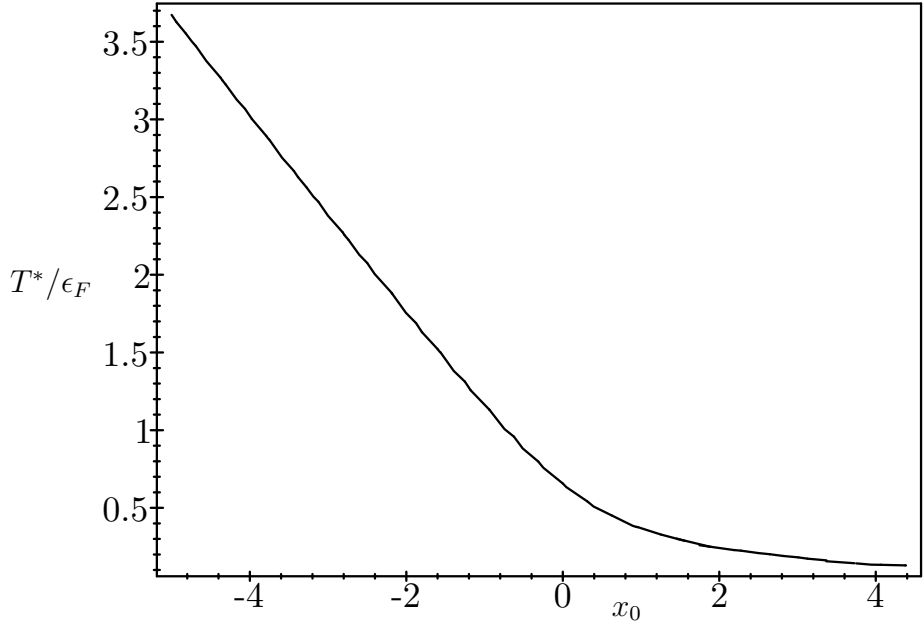
%##
\input 3d.tps  \\[-2cm]
\caption{
Dependence of $T^*$
on crossover parameter
in
three dimensions.}
\label{f5}\end{figure}

\begin{figure}[tbh]
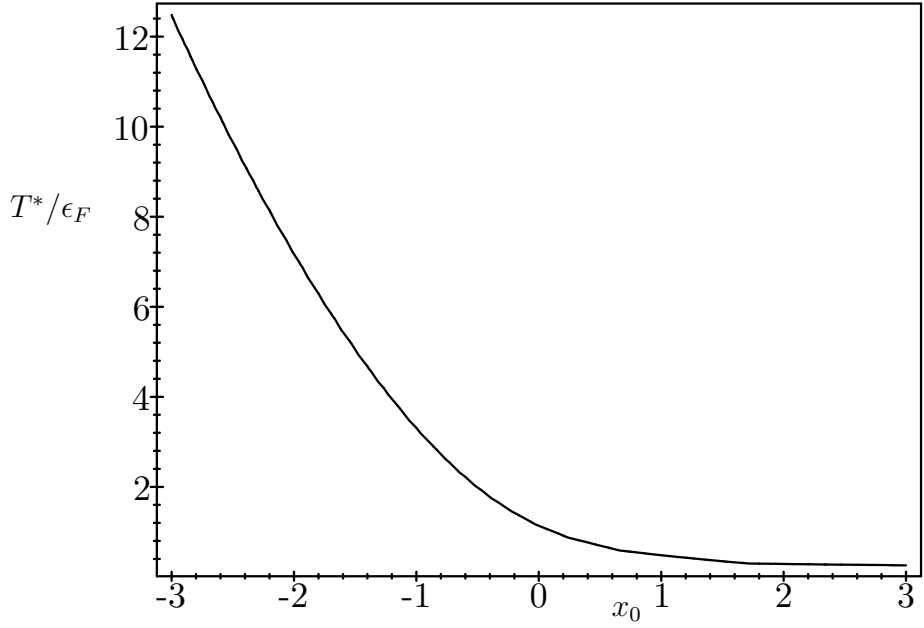
%##
\input 2d.tps  \\[-2cm]
\caption{
Dependence of $T^*$
on crossover parameter
in
two dimensions.}
\label{f6}\end{figure}

Let us now turn to the region near zero temperature,
where we
can derive exact results for
gap.
 From (\ref{1.6}) we extract the asymptotic behavior
in the three-dimensional case for $x_0>1$.
In this region one can assume density of states to be  roughly constant,
since the integrand of (\ref{1.3.1})
is  peaked
in the narrow region near $\xi=0$. The small-$T$ behavior is
\begin{eqnarray}
\Delta (T) = \Delta (0)  - \Delta (0) \sqrt{ {\pi \over 2} }
\sqrt{ { T \over \Delta (0) } }
\exp\left[- {  \Delta (0) \over T } \right]  \left[  1+ {\rm erf}
\left( \sqrt{\frac{\sqrt{x_0^2+1}-1}{T/ \Delta(0)}}
%{ \mu \over \Delta (0)} \sqrt{ { \Delta (0) \over 2T } }
 \right) \right],
\label{3d0}
\end{eqnarray}
where ${\rm erf}(x)$ is the error function.
Since the density of states
is nearly constant in this limit,
the same equation  holds in two-dimensions---apart from  a modified gap
 $\Delta (0) $ given by (\ref{1.13}).

In the weak-coupling limit,
 $x_0=\lfrac{ \mu}{ \Delta (0) } $ tends to infinity,
and
the expression above approaches exponentially fast
the well-known BCS result:
\begin{eqnarray}
\Delta(T)
=\Delta(0) - [2 \pi \Delta(0) T] ^{1/2}\exp[-{ \Delta(0) \over T } ]
\label{2dbcs}
\end{eqnarray}
For strong couplings with
 $ x_0 < - 1 $, the three-dimensional
 integrands are
no longer peaked in the narrow region
so that the density of states can no longer be taken to be constant.
Taking this into  account, we find:
\begin{eqnarray}
\Delta(T) = \Delta(0) - { 8 \over \sqrt{\pi} } \sqrt{-x_0}
\left( \Delta(0) \over T \right)^{3/2} \exp
\left[ - \frac{\sqrt{\mu^2+\Delta^2(0)}}{T}
\right].
 \label{3dbe}
 \end{eqnarray}
 From Eq.~(\ref{3dbe}) we see that near $T=0$ the gap
$\Delta(T) $ tends
in the strong-coupling limit
 exponentially to $ \Delta(0) $,
forming plateau near $ T=0$.

In two dimensions
we arrive at similar result: an
exponentially growing plateau near $T=0$ in the
strong  coupling  limit:
\begin{eqnarray}
\Delta(T) = \Delta(0) - { \Delta(0) \over 2 }  E_1 \left(
{ \sqrt{\Delta(0)^2+\mu^2}  \over T } \right),
\label{2dbe}
\end{eqnarray}
where $E_1$ is the    exponential integral
$E_1 (z)= \int_z^{\infty} e^{-t}/t~ dt $.

For very strong couplings, Eq.~(\ref{2dbe}) becomes:
\begin{eqnarray}
\Delta (T) = \Delta (0)- { \Delta (0) \over 2 }
{T \over \sqrt{\mu^2+\Delta^2(0)} }
{\rm exp} \left( - { \sqrt{\mu^2 + \Delta(0)^2} \over  T} \right)
 \label{extr}
 \end{eqnarray}

Let us also calculate thermodynamical quantities
near $T=0$.
For the thermodynamic Gibbs potential
$ \Omega (T,\mu, V)$  we calculate
\begin{eqnarray}
 \Omega  = \sum_{\bf k} \left\{
{ \Delta ^2 \over 2 \sqrt{\xi^2_{\bf k} + \Delta^2 }  }
 \tanh  { \sqrt{\xi^2_{\bf k} + \Delta^2} \over 2T }  -
2 T \log \left[ 2 \cosh   { \sqrt{\xi^2 _{\bf k} + \Delta^2} \over 2 T }\right]
  +\xi_{\bf  k}
\right\}.
\label{3om}
\end{eqnarray}
Here and in the sequel in this section, $ \Delta(0) $ will be replaced by
$ \Delta $.
In three dimensions, Eq.~(\ref{3om}) turns into the
\begin{eqnarray}
\!\!\!\!\!\!\!\!\!{ \Omega  \over V }
 &\!\! = \!\!& \kappa_3 \int_{-\mu}^{\infty}  d\xi
 \sqrt{\xi + \mu}
\left[
{ \Delta ^2 \over 2 \sqrt{\xi^2 + \Delta^2 }  }
 \tanh
{  \sqrt{\xi^2 + \Delta^2} \over 2T }
 -
2 T \log \left( 2 \cosh
  {  \sqrt{\xi^2 + \Delta^2} \over 2 T } \right)
  +\xi
\right]\!,
\label{i3om}      \nonumber \\&&
\end{eqnarray}
In two dimensions, we obtain instead:
\begin{eqnarray}
{ \Omega  \over V }
=\kappa_2 \int_{-\mu}^{\infty} d\xi
\left[
{ \Delta ^2 \over 2 \sqrt{\xi^2 + \Delta^2 }  }
 \tanh
{  \sqrt{\xi^2 + \Delta^2} \over 2T }
-
2 T \log \left( 2 \cosh
  {  \sqrt{\xi^2 + \Delta^2} \over 2 T }
 \right)  +\xi
\right],
\label{i2om}
\end{eqnarray}
We regularize the thermodynamic potential $\Omega_s $
of the  condensate
subtracting $ \Omega_n= \Omega (\Delta = 0)$.
 At $T=0$ and  for weak couplings  this is found to depend
 on temperature as follows:
\begin{eqnarray}
 \frac{ \Omega _s}{V} \equiv { \Omega - \Omega_n \over V} =
\kappa_3 \sqrt{\mu}  \left[ - { \Delta^2\over  4}
+{1 \over 2}  \mu | \mu | - { 1 \over 2} \mu \sqrt{\mu^2 +\Delta^2 }
  \right]
 \label{o3t0}
\end{eqnarray}
In the BCS limit ($x_0 \rightarrow \infty$) this reduces
to  the well-known  result
\begin{eqnarray}
{ \Omega_s  \over V} =
\kappa_3 \sqrt{\mu}  \left[ - { \Delta^2\over  2}
 \right].
\label{o3t0scs}
\end{eqnarray}
In two dimensions, we have a formula
valid for any strength of coupling:
\begin{eqnarray}
{ \Omega _s \over V} =
\kappa_2  \left[ - { \Delta^2\over  4}
+{1 \over 2}  \mu | \mu | - { 1 \over 2} \mu \sqrt{\mu^2 +\Delta^2 }
 \right],
 \label{o2t0}
\end{eqnarray}
with
the BCS limit
\begin{eqnarray}
{ \Omega_s  \over V} =
\kappa_2 \left[ - { \Delta^2\over  2}
 \right].
\label{o2t0scs}
\end{eqnarray}
In both three- and two-dimensional cases in the BCS limit
we can write a small temperature correction ${ \pi T^2 / 3 }$
to the thermodynamic potential.
In the opposite limit of strong couplings,
 we find
in three dimensions the strong-coupling limit:
\begin{eqnarray}
 {\Omega \over V}  =
- {\pi \over 64 } \kappa_3  \Delta^{5/2}  (-x_0)^{-3/2}.
\label{o3dstr}
\end{eqnarray}
The gap $ \Delta (0)$ has by Eq.~(11) the strong-coupling
limit $ \Delta (0) \approx \epsilon_F [16/3 \pi ]^{3/2}|x_0|^{1/3}$
yielding the large-$x_0$ behavior

\begin{equation}
\frac{ \Omega}{V}  \sim  - \kappa_3 \epsilon_F^{5/2}
\frac{\pi}{64} \left(
\frac{16}{3\pi} \right)^{15/4}
 |x_0|^{-2/3}.
\end{equation}
In two dimensions, we substitute
 the gap
function
$\Delta$ of Eq.~(\ref{1.13}),
into the
 thermodynamic potential (\ref{o2t0}), and  obtain  for
 strong couplings where
 $ \mu < 0$:
\begin{eqnarray}
\frac{ \Omega}{V}  \equiv 0
 \label{o2t0s}
\end{eqnarray}

Let us now turn to the entropy.
In three dimensions  near $T=0$ it is given for weak couplings  by:
\begin{eqnarray}
         { S \over V } =
         \kappa_3
        \sqrt{\mu}
\left\{
        \sqrt{ {  2 \pi \Delta ^3 \over T} }\exp
        \left(-
        {  \Delta \over T}
        \right)
\left[  1+ {\rm  erf}
        \left(
      \sqrt{  {\sqrt{ x_0^2+1}-1  \over T / \Delta   }}
\right)
\right]
 + 2 \mu
        \exp
        \left( -
        {  \sqrt{\mu^2 + \Delta^2} \over T }
        \right)
\right\},
\label{s}
\end{eqnarray}
For $ \mu / \Delta \rightarrow \infty$,
this reduces correctly  to the  BCS result:
\begin{eqnarray}
\frac{S}{V} =   \kappa_3  \sqrt{\mu}  \sqrt{ {  8 \pi \Delta ^3 \over T} }
\exp \left( -  { \Delta \over T} \right)
\label{sbcs}
\end{eqnarray}
In  two dimensions,
the  result is similar [with $ \Delta=\Delta(0)$ given by Eq.~(\ref{1.13})]:
\begin{eqnarray}
 { S \over V  } =   \kappa_2 \left\{ \sqrt{ {  2 \pi \Delta ^3 \over T} }
\exp \left( - { \Delta \over T} \right)
    \left[  1+ {\rm erf}  \left(
 \sqrt{  {\sqrt{ x_0^2+1}-1  \over T / \Delta   }}
% { \mu \over \sqrt{2 \Delta  T}  }
 \right)\right]
+ 2 \mu
\exp \left( - { \sqrt{\mu^2 + \Delta^2} \over T }
   \right)  \right\},
\label{s2d}
\end{eqnarray}
The BCS limit of this is
\begin{eqnarray}
\frac{S}{V} =   \kappa_2 \sqrt{ {  8 \pi \Delta ^3 \over T} }
\exp \left( -  { \Delta \over T} \right)
\label{sbcs2}
\end{eqnarray}
In the strong-coupling limit
where  $ \mu/ \Delta \ll -1$,  we have for the entropy
in three dimensions
\begin{eqnarray}
\frac{S}{V}=  \kappa_3  { \sqrt{\pi} \over 4 }  T^{1/2 }
 \sqrt{\mu^2+\Delta^2}
  \exp\left( -
\frac{
  \sqrt{\mu^2+\Delta^2}}{T}
\right)
,
 \label{l1}
\end{eqnarray}
and in two dimensions:
\begin{eqnarray}
\frac{S}{V}=  - 2  \kappa_2 \mu
\exp\left( -\frac{  \sqrt{\mu^2+\Delta^2}}{ T}
\right)
\label{l2}
\end{eqnarray}
 From the entropy, we  easily derive
the
heat capacity at a constant volume
$c_V$.
In three dimensions it is given
near $T=0$  for weak
couplings by
\begin{eqnarray}
c_V=  \kappa_3 \sqrt{\mu}  \sqrt{2 \pi \Delta ^3} \left\{  { \Delta \over
T^{3/2} }
\exp \left( -{  \Delta \over T} \right) \left[ 1+ {\rm erf} \left(  \sqrt{
\frac{ \sqrt{x_0^2+1} -1  }{T/\Delta}} \right)
\right]
\right\}
\label{c3d}
\end{eqnarray}
reducing in the limit
 ${x_0 \rightarrow \infty }$
to the BCS result
\begin{eqnarray}
c_V =  \kappa_3 \sqrt{\mu}  \sqrt{2 \pi \Delta ^3} { 2\Delta \over T^{3/2} }
\exp \left(- {  \Delta \over T}  \right).
\label{c3dbcs}
\end{eqnarray}
In two dimensions, the weak-coupling behavior is
\begin{eqnarray}
c_V =  \kappa_2 \sqrt{2 \pi \Delta ^3} \left\{  { \Delta \over T^{3/2} }
\exp\left( - {  \Delta \over T} \right)  \left[ 1+
{\rm erf} \left(  \sqrt{
\frac{ \sqrt{x_0^2+1} -1  }{T/\Delta}} \right)
%{\rm erf} \left(  { \mu\over \sqrt{2 \Delta T}  } \right)
\right]
\right\},
\label{c2d}
\end{eqnarray}
while  the strong-coupling behavior in three dimensions is
\begin{eqnarray}
c_V =
\kappa_3  { \sqrt{\pi} \over 4 }  T^{-1/2 }
 (\mu^2+\Delta^2)
 \exp\left( -
\frac{
  \sqrt{\mu^2+\Delta^2}}{T}
\right)
%  \kappa_3   { \sqrt{\pi} \over 4 }
%{ \mu^2 \over T^{1/2} } e^{\mu /T},
\label{cstr3d}
\end{eqnarray}
and in two dimensions
\begin{eqnarray}
c_V = 2  \kappa_2   {\mu^2 \over T } \exp \left(
-\frac{\sqrt{\mu^2+\Delta^2}}{T}
\right).
\label{cstr2}
\end{eqnarray}

\section{Crossover from  BCS Superconductivity Near $T_c$
 to Onset of  Pseudogap Behavior}

We now turn to the region near $T^*$,
for which we  derive asymptotic behavior of
the  ratios $\lfrac{ \Delta (T)}{ T^* } $
and  $\lfrac{ \Delta (T)}{ \Delta(0) } $
as well as thermodynamic quantities.
In doing so, we shall consider $\Delta (T) /T$
as a small parameter of the problem. In the calculations near $T^* $
it is convenient to use $ \lfrac{ \mu}{ 2 T^*  } $
(this ratio tend to $\infty$ in the weak-coupling limit
and to the $-\infty$ in the strong coupling one)
as a crossover parameter rather than $x_0$.
In three dimensions, we find  for weak couplings
\begin{eqnarray}
 \left[
        { \Delta (T) \over  2T^*}
 \right]^2 =
{ \displaystyle
        \left( 1- { T \over T^* } \right)
 \left(
        1+ \tanh
        { \mu \over 2 T^*  }
 \right)
\over
    \displaystyle
        { 1 \over 4 }
 \left[ \displaystyle
          {1 \over \mu /2T^*  } -
        {  \displaystyle
 1 \over (\mu /2 T^* )^2 }
        \tanh  { \mu \over 2 T^*  }
        \displaystyle
 \right] +
  \displaystyle
 \left(
        { 2 \over  \pi}
 \right) ^2 \
 \left(
        1 + \displaystyle
 { 2 \over \pi }
       \arctan
  \displaystyle
 \frac{\mu }{ \pi T^*}
 \right)
}.
\label{1tc1}
\end{eqnarray}
In the limit $ \lfrac {\mu }{ 2 T^*}  \rightarrow \infty$
this tends to the BCS result
\begin{eqnarray}
{\Delta(T) \over T_c} \simeq  3 \sqrt{1 - { T\over T_c} }.
\label{tc1bcs}
\end{eqnarray}
In the opposite limit of strong couplings,
 $T^* $ and $ \Delta (0)$ tend to infinity. The ratio
$ \Delta (T) /T $ near $T^* $
 tends to zero exponentially as a function of the
crossover parameter $\mu / 2 T^*$:
\begin{eqnarray}
\left[
{ \Delta (T) \over 2 T^* }
\right]^2
 = { 16 \over  \sqrt{2 \pi} }
\left(
1 - { T \over T^* }
\right)
\left( -
{ \mu \over 2 T^* }
\right)^{3/2}
 e^{\mu /T^*}.
\label{tcstr3d}
\end{eqnarray}
In two-dimensions the  near-$T^* $
formula (\ref{1tc1}) holds over the crossover region:
In the weak-coupling limit,
this formula
 reproduces the BCS result (\ref{tc1bcs}).
In the strong-coupling limit, we find  as in three dimensions a
 ratio  $ \Delta (T) /T $ which tends
to zero exponentially
as a function of the crossover parameter $\mu/T^*$ :
\begin{eqnarray}
 \left[
{ \Delta (T) \over  2T^*}
 \right]^2 =
2 \left[ { 1 \over 4 } - \left(  { 2 \over  \pi} \right)^2   \right] ^{-1}
\left(  { \mu \over  2 T^* } \right)
\left( 1 - { T \over  T^* } \right)
e^{\mu /T}.
\label{n1}
\end{eqnarray}

Let us calculate
the
dependence of  $T^* $
on
the crossover parameter $ {\mu/ 2 T^*}$ in the strong-coupling
limit.
In
three dimensions, we obtain from
Eq.~(\ref{1.3.1})
the relation
\begin{eqnarray}
{ T^* \over \epsilon _F} =
\left(
{ 1 \over 3} \right)^{2/3}
  \exp \left( - { 2 \over 3 }  { \mu \over  T^*  } \right).
\label{n2}
\end{eqnarray}
This is solved for $T^*$
(up to a logarithm) by
\begin{eqnarray}
T^* \simeq - \frac{2}{3} \mu \log^{-1} \left( - \frac{\mu}{\epsilon_F} \right)
\end{eqnarray}
[see also the discussion after the formula (\ref{1.5})].
As a function of
the
crossover parameter $x_0$ we obtain
\begin{eqnarray}
{T^* \over \epsilon_F} \simeq {1 \over 2}
\left( { 16 \over 3 \pi}
\right)^{2/3} |x_0|^{4/3} \log^{-1}
\left( \sqrt{16/ \pi} |x_0|
\right).
\end{eqnarray}
In two dimensions we find
from (\ref{1.3.2})
\begin{eqnarray}
{ T^* \over \epsilon _F} =
{ 1 \over 2 } \exp \left(  -   { \mu  \over  T^* }  \right).
\label{n3}
\end{eqnarray}
and thus
\begin{eqnarray}
T^* \simeq - \mu \log^{-1} \left( - \frac{\mu}{\epsilon_F} \right)
\end{eqnarray}
As a function of $x_0$, this implies
\begin{eqnarray}
{T^* \over \epsilon_F}=
2 x_0^2
 \log^{-1}
\left( 2 \sqrt{2} |x_0|
\right).
\label{a01}
\end{eqnarray}
Let us  also derive in the strong-coupling region  the dependence
of the ratio $  \lfrac{\Delta (0) }{  T^* }$
on the crossover parameter which in three dimensions reads
\begin{eqnarray}
{ \Delta (0) \over  T^*  } =
{ 4 \over \sqrt{\pi} }
\left( - {\mu  \over T^*  } \right) ^{1/4}
\exp \left(   {  \mu \over 2 T^*  } \right),
\label{n4}
\end{eqnarray}
and in two dimensions:
\begin{eqnarray}
{ \Delta (0) \over  T^*  } =
4  \left( -  {\mu  \over 2 T^*  } \right) ^{1/2}
\exp \left(   {  \mu \over 2 T^*  } \right).
\label{4n4}
\end{eqnarray}
In the weak-coupling  regime, both three- and two- dimensional cases
yield the result
\begin{eqnarray}
{ \Delta (0) \over T^* } = { \pi \over e^\gamma }
\left(1-{\Delta (0)^2 \over 4 \mu^2} \right)^{-1/2}=
 { \pi \over e^\gamma }
\left(1-{1 \over 4 x_0^2 }\right)^{-1/2}
\simeq
 { \pi \over e^\gamma }
\left(1+{1 \over 8 x_0^2 }\right)
{}.
\label{n5}
\end{eqnarray}
The temperature $T^*$
is in the weak-coupling regime of three- and
two dimensional systems
the following function of $x_0$:
\begin{equation}
\frac{T^*}{\epsilon_F} \simeq
\frac{ e^\gamma}{ \pi  }
\left( \frac{1}{x_0} -{3 \over 8 x_0^3 }\right)
\end{equation}

Using these results,
 we can also calculate the
asymptotic behavior of the ratio
 $ \lfrac{ \Delta  (T) }{  \Delta (0)} $  near $T^*$.
In three dimensions,  the strong-coupling limit yields
\begin{eqnarray}
\left[
 \displaystyle{ \Delta (T) \over \Delta (0) }
\right]^2
=
{ \sqrt{ \pi} \over 2 }
 \left( - { \mu \over 2 T^* }  \right)
 \displaystyle\left( 1 - { T \over T^* } \right).
\label{n6}
\end{eqnarray}
In two dimensions we have that in the strong-coupling limit
this ratio tends to
\begin{eqnarray}
\left[
{ \Delta (T) \over \Delta (0) }
\right]^2
= { 1 \over 8}
\left(
{ 4 \over \pi ^2 }-
{ 1 \over 4 }
\right) ^{-1}
\left( 1 - { T \over T^*  } \right).
\label{n7}
\end{eqnarray}
At weak couplings, both three- and two-dimensional gap functions
are given by
\begin{eqnarray}
 \left[
        { \Delta (T) \over  \Delta (0)}
 \right]^2 =
{ 4 \pi ^2 \over e^{2\gamma} }
{\displaystyle
        \left(1- { T \over T^* } \right)
 \left[
        1+ \tanh
        { \mu \over 2 T^*  }
 \right]
\over
     \displaystyle      { 1 \over 4 }
 \left[
          {1 \over \mu /2T^*  } -
        { 1 \over (\mu /2 T^* )^2 }
        \tanh
        {\mu \over 2 T^*}
 \right] +
\displaystyle \left(
        { 2 \over  \pi}
 \right) ^2 \
 \left(
        1 + { 2 \over \pi }
        \arctan
        { \mu  \over \pi T^* }
 \right)
}
\label{n8}
\end{eqnarray}
In order to calculate thermodynamic potential near $T^*$
we expand the general expression by $ \lfrac{\Delta (T)}{ \Delta(0) }$
and keeping the terms of the same order we get :
\begin{eqnarray}
\!\!\!\!\!\!\frac{\Omega_s}{V} & \simeq &
  -
{ (T^* - T) \Delta^2 \over  4 T^* }
\int
{ d^D {\bf p} \over  (2 \pi)^D}
\cosh^{-2}   {  \xi \over  2 T^* }
\nonumber  \\
&& - {\Delta^4 \over 8}
\int
{ d^D {\bf p} \over  (2 \pi)^D}
{ 1 \over  \xi^2}
\left(  { 1 \over 2 T^*  }
\cosh^{-2}   {  \xi \over  2 T^* }
- { 1 \over \xi} \tanh { \xi \over  2T^* }
\right)
\label{1n1}
\end{eqnarray}
where $D$ is the space dimension.
  Note once more
 that
  we consider here the temperature evolution of the system
at a fixed chemical potential ($ \mu(T,1/k_F a_s) = \mu(0, 1/k_F a_s)$)
and regularize $\Omega$ by a subtraction of  $\Omega_n = \Omega (\Delta=0)$.

The thermodynamic potential in three dimensions in the weak-coupling regime
 near $T^* $ is given
by
\begin{eqnarray}
\label{om3Dtc}
&& \!\!\!\!\!\!\!\!\!   \frac{ \Omega _s}{V} =
- \kappa_3 \sqrt{\mu }
\Biggl\{
        { (T^* -T) \Delta ^2 \over  2 T^* }
        \left[
        1+ \tanh
        {\mu \over 2 T^*}
        \right]+
 \\
&&     ~~~~     + {\Delta ^4 \over 4}
        { 1 \over (2 T^* )^2 }
\left[
        { 1 \over 4 }
        \left(
         {1 \over \mu /2T^*  }
        - { 1 \over (\mu /2 T^* )^2}
        \tanh
        {\mu \over 2 T^*}
        \right) +
        \left(
        { 2 \over  \pi}
         \right)  ^2
        \left(
        1 + { 2 \over \pi }
        \arctan
        \frac{\mu }{ \pi T^*}
        \right)
\right] \Biggr\}.
 \nonumber
\end{eqnarray}
In the  BCS limit, this reduces to the well-known formula:
\begin{eqnarray}
\frac{ \Omega _s}{V}
=  - \kappa_3 \sqrt{\mu}
\Delta ^2
\left(  1- { T \over  T_c } - { 1 \over 2 \pi ^2 }  { \Delta ^2  \over T_c ^2 }
 \right)
\label{ombc}
\end{eqnarray}
In the strong-coupling limit we have:
\begin{eqnarray}
         {\Omega_s \over V } =
       - \kappa_3
\left\{
        { \pi \over 64 }
         \Delta ^4 (2 T^*)^{-3/2}
        \left(
        - { \mu \over 2T^*}
        \right)
        ^{-3/2}  +
        \left(
        1- { T\over T^* }
        \right)
\Delta ^2
        {\sqrt{ \pi}  \over  2}
        \sqrt{T^* } \exp
        \left(
        { \mu \over  T^* }
        \right)
\right\}
\label{strtc3d}
\end{eqnarray}
 Using the asymptotic estimates
 derived  above for  the
strong-coupling limit,
  and the fact that in this limit
\begin{eqnarray}
  \frac{\mu }{T^*} \simeq - \log \left(-\frac{\mu }{\varepsilon_F} \right)
\simeq - { \rm const} \times \log (|x_0|),
\label{td2}
\end{eqnarray}
we find  near $ T^* $
 the  difference between
the thermodynamic potential of the gapless
and pseudogapped normal states:
\begin{eqnarray}
{\Omega_s \over V} \simeq
- {\rm const}
\left(
        1- { T\over T^* }
\right) ^2
| x_0|^{-3/2}.
\label{1n2}
\end{eqnarray}
In two dimensions near $T=T^* $, the thermodynamic potential
of the gas of pairs
is given by the formula holding for the crossover region
\begin{eqnarray} \label{om2Dtc}
   && \!\!\!\!\!\!\!\!   {\Omega_s \over V} =
- \kappa_{2}
\Biggl\{
        { (T^* -T) \Delta ^2 \over  2 T^* }
        \left[
        1+ \tanh
        {\mu \over 2 T^*}
        \right]+
 \\
   &&  ~~~~      + {\Delta ^4 \over 4}
        { 1 \over (2 T^* )^2 }
\left[
        { 1 \over 4 }
        \left(
         {1 \over \mu /2T^*  }
        - { 1 \over (\mu /2 T^* )^2}
        \tanh
        {\mu \over 2 T^*}
        \right) +
        \left(
        { 2 \over  \pi}
         \right)  ^2
        \left(
        1 + { 2 \over \pi }
        \arctan
        \frac{\mu }{ \pi T^*}
        \right)
\right]
\Biggr\}.
 \nonumber
\end{eqnarray}
In the BCS limit, this yields the familiar result
 \begin{eqnarray}
\frac{\Omega_s}{V}
=  - \kappa_2
\Delta ^2
\left(  1- { T \over  T^* } - { 1 \over 2 \pi ^2 }  { \Delta ^2  \over T^{* \
2} }
 \right),
\label{m1}
\end{eqnarray}
and in the
strong-coupling limit:
\begin{eqnarray}
 \frac{ \Omega _s}{V} =
- \kappa_2 \left\{
\left( 1- { T \over T^* }   \right)
\Delta^2 \exp \left(  { \mu \over  T^* } \right)
+{ \Delta ^4  \over  4}
{ 1 \over (2 T^*)^2 }
\left[   \left(
{ 1 \over 4} -
 { 4 \over  \pi^2}
\right)
{1 \over  \mu /2 T^* }
\right]
 \right\} .
\label{extrsc}
\end{eqnarray}
Using the earlier-derived asymptotic
behavior plus the  limiting equation (\ref{td2}) which also
holds for two-dimensional case,
we derive for the thermodynamic potential the
 $x_0$-behavior
\begin{eqnarray}
{ \Omega _s \over V} \simeq
- {\rm const} \times \left(  1 - { T \over T^* } \right) ^2
\log |x_0|.
\label{12n1}
\end{eqnarray}
The entropy behaves near $T^* $ in three dimensions in the weak-coupling
regime like
\begin{eqnarray}
\frac{S_s}{V} \equiv  \frac{S-S_n}{V} =
- \kappa_3 \sqrt{\mu} {\Delta ^2   \over 2 T^*  }
\left[
1+ \tanh \left(  { \mu \over 2 T^*  } \right)
\right]
\label{s3dtc}
 \end{eqnarray}
with the BCS limit
\begin{eqnarray}
\frac{S_s}{V} =
-   \kappa_3 \sqrt{\mu} {\Delta ^2   \over  T_c  }.
\label{s3tcbcs}
\end{eqnarray}
The opposite strong-coupling limit is in
three dimensions:
\begin{eqnarray}
\frac{S_s}{V} =
-\kappa_3   { \sqrt{ \pi}  \over 2}
\Delta ^2 {T^*}^{-1/2}
\exp \left( { \mu \over  T^* } \right).
\label{sbetc}
\end{eqnarray}
Inserting the  above asymptotic formulas
for $\Delta , ~\mu , ~ T^* $, we find
\begin{eqnarray}
\frac{S_s}{V} \simeq - {\rm const} \times
\left( 1 - { T  \over T^*   } \right)
| x_0| ^{-5/3} .
\label{ns1}
\end{eqnarray}
In two dimensions,  the entropy  is given in
 the entire crossover region by
\begin{eqnarray}
 \frac{S_s}{V} =
-  \kappa_2  {\Delta ^2   \over 2 T^*  }
\left[
1+ \tanh   { \mu \over 2 T^*  }
\right]                ,
\label{s2dtc}
 \end{eqnarray}
and has the  BCS limit
\begin{eqnarray}
\frac{S_s}{V}  =
- \kappa_2  {\Delta ^2   \over  T^*  } ,
\label{s2tcbcs}
\end{eqnarray}
while  the strong-coupling limit  becomes:
\begin{eqnarray}
\frac{S_s}{V} =
-  \kappa_2  {\Delta ^2   \over  T^*  }
 e^{\mu /T^*}.
\label{s2dtcsc}
 \end{eqnarray}
Using corresponding asymptotic formulas for
 $ \Delta ,~\mu ,~T^*$
 in two dimensions, this depends on  $x_0$
as
\begin{eqnarray}
\frac{S_s}{V} = -
{\rm const} \times
\left( 1 - { T  \over T^*   } \right)
 x_0^{-2}                 .
\label{ns2}
\end{eqnarray}
In order
to derive the specific heat we must take into  account the
temperature dependence of the gap.

In three dimensions, we find in the
 weak-coupling region near $T^*$:
\begin{eqnarray}
 \frac{C_s}{V}  =
      2T  \kappa_3 \sqrt{ \mu}
{\displaystyle
 \left(
        1+ \tanh
        { \mu \over 2 T^*  }
 \right)^2
\over
    \displaystyle
  {  1 \over 4 }
 \left[
          {1 \over \mu /2T^*  } -
        { 1 \over (\mu /2 T^* )^2 }
        \tanh
        {\mu \over 2 T^*}
 \right] +
 \left(
        { 2 \over  \pi}
 \right) ^2 \
 \left(
        1 + { 2 \over \pi }
        \arctan
    \frac{\mu }{ \pi T^*}
\right)
},
\label{heattc3d}
\end{eqnarray}
which has the well-known BCS limit:
\begin{eqnarray}
 \frac{C_s}{V}  \simeq
  \kappa_3  \sqrt{ \mu} \pi^2 T_c .
\label{heatbcs}
\end{eqnarray}
In the strong-coupling limit, we find in three
dimensions
\begin{eqnarray}
{ C_s \over V} = \kappa_3
16 \sqrt{2} {T^*} ^{3/2}
\left( - { \mu \over 2T^* } \right)^{3/2}
  e^{2\mu /T^*}
\label{heat3be}
\end{eqnarray}
 Inserting earlier
 derived  asymptotic formulas
 we see that $C_s $
tends in the strong-coupling limit to zero like
\begin{eqnarray}
 \frac{C_s}{V} \sim  {\rm const}\times |x_0|^{-2}
\label{n1c}
\end{eqnarray}
In two dimensions, the result
 for the entire crossover region reads
\begin{eqnarray}
{ C_s \over V} =
 2 T^*  \kappa_2
{     \displaystyle
 \left(
        1+\tanh
        { \mu \over 2 T^*  }
 \right)^2
\over   \displaystyle
              { 1 \over 4 }
 \left[
          {1 \over \mu /2T^*  } -
        { 1 \over (\mu /2 T^* )^2 }
        \tanh
        {\mu \over 2 T^*}
 \right] +
 \left(
        { 2 \over  \pi}
 \right) ^2 \
 \left(
        1 + { 2 \over \pi }
        \arctan
 \frac{\mu }{ \pi T^*}
\right)
}.
\label{2heattc3d}
\end{eqnarray}
This becomes in the BCS limit
\begin{eqnarray}
 \frac{C_s}{V} \simeq
   \kappa_2 \pi^2 T^*,
\label{2dheatbcs}
\end{eqnarray}
and in the  strong-coupling limit
\begin{eqnarray}
\frac{C_s}{V} =
 4 \kappa_2
\mu \left( { 1 \over 4} -
{4 \over \pi^2} \right )^{-1}
\exp \left(  { 2 \mu \over T^*}\right).
\label{n2c}
\end{eqnarray}
As a function of $x_0$, the result is
\begin{eqnarray}
\frac{C_s}{V} \sim {\rm const} \times x_0^{-2}.
\label{n3c}
\end{eqnarray}
{}From the above calculation near $T^*$
we see that both
  quantities $S_s$ and $C_s$
tend quickly to zero with growing coupling
strength
in the pseudogapped regime
(like a power of the
crossover parameter  $|x_0|$ or exponentially as
a function of crossover parameter $\lfrac{\mu }{ 2 T^*}$).
So, at very strong couplings
$T^*$ is getting less and less pronounced
with increasing coupling strength.

Note that
in the strong-coupling regime,
the modified gap function $\sqrt{\mu^2+\Delta^2}$
 [see  Eq.~\ref{1.5.1}]  enters the
expressions for thermodynamical quantities below $ T^*$
the same way
as an ordinary gap
in BCS limit
 [see Eqs.~(\ref{l1}), (\ref{l2}), (\ref{cstr3d}), (\ref{cstr2})].

\section[Phase fluctuations in Two Dimensions and Kosterlitz-Thouless
Transition]
{Phase fluctuations in Two Dimensions \\ and Kosterlitz-Thouless Transition}

In the previous sections we have calculated the
properties of the model in the
mean-field
 approximation.

Now we are ready to go beyond the mean-field
approximation, which
as discussed in
the introduction, supply us with modulus of
the gap function.
In this chapter we make use of derivative
expansion which determines
 the crucial
stiffness parameter
for the study of
phase fluctuations, that in two dimensions
leads to the Kosterlitz-Thouless transition,
at which the
expectation of the complex order field $ \Delta({\bf x})=|\Delta|({\bf x}) e^{i
\theta({\sbf x})}$ vanishes
in the pseudogap state.
In these calculations we
assume with other authors
 (\cite{shar}, \cite{Po})
 that
the phase fluctuations do not significantly affect
modulus of $\Delta$.
We shall first study
the
two-dimensional system,
where
the mean-field solution receives
the strongest
modifications
from the violent phase fluctuations,
as articulated
by the
 Coleman-Mermin-Wagner-Hohenberg theorem
 \cite{col} which forbids
the existence of a strict long-range order,
leading to a power behavior of
correlation functions for all temperatures below $T_{KT}$.

%According to Ref.~\cite{Em}, a
%  Kosterlitz-Thouless transition can give an adequate description of the
% superconductive transition in underdoped cuprates.
The crossover of the  Kosterlitz-Thouless transition from weak to strong
coupling was first considered in \cite{Em,Dr}, and studied
 recently by means of an XY-model
in \cite{shar}, with
the stiffness derived
from
a fixed
nonvanishing modulus of the order parameter $ \Delta $.
In Appendix A, we outline the derivation of the
effective Hamiltonian in \cite{shar}.
Under the same assumptions, we shall analyze
the
weak- to strong-coupling
crossover of the  Kosterlitz-Thouless transition in the present work.

\old{
{\tt In this section we introduce
modulus-phase variables and replace
$\xi_\sigma (x) = \hi_\sigma (x) exp (i \theta (x) /2)$
physical meaning of such replacing is equal
to introducing neutral fermion $\xi_\sigma (x)$
and charged boson $exp( i \theta (x) /2 )$.
Persistence of the gap neutral fermions
in the spectrum of charged fermions $\xi$
even above $T_{KT}$
in two-dimensional case
was discussed in \cite{W}.}}

Writing the spacetime-dependent order parameter
as $ \Delta(x)e^{i\theta({x})}$,
where $x$ denotes the four-vector $x=(\tau ,{\bf x})$
formed from imaginary time and position vector,
the partition function
may be written as a functional integral
\cite{6',shak,W,T}
\begin{equation} Z(\mu, T) = \int \Delta \,{\cal D} \Delta\,
{\cal D} \theta \exp{[-\beta \Omega (\mu, T, \Delta(x), \partial \theta
(x))]},
\label{bk1}
\end{equation}
where
\begin{equation}
\beta \Omega ( \mu , T, \Delta (x), \partial \theta (x)) =
\frac{1}{g} \int_{0}^{\beta}
d \tau \int d {\bf x} \Delta^{2}(x) -
\mbox{Tr log} G^{-1} + \mbox{Tr log} G_{0}^{-1}
\label{dfg}
\end{equation}
is  the one-loop effective action,
containing the
inverse Green function of the fermions
in the collective pair field
\begin{eqnarray}
G^{-1} & = & -\hat{I} \partial_{\tau} +
\tau_{3} \left(\frac{\nabla^{2}}{2 m} + \mu \right) +
\tau_{1} \Delta(\tau, \mbox{\bf x})
\nonumber \\
& - &
\tau_{3} \left[ \frac {i \partial_{\tau} \theta(\tau, \mbox{\bf x})}{2} +
\frac{(\nabla \theta(\tau, \mbox{\bf x}))^{2}}{8m} \right] +
\hat{I} \left[\frac{i \nabla^{2} \theta(\tau, \mbox{\bf x})}{4 m} +
\frac{i \nabla \theta(\tau, \mbox{\bf x}) \nabla}{2m} \right].
\label{bk3}
\end{eqnarray}
Here $\tau_1 ,~ \tau_3$ are the usual Pauli matrices,
and
$G_{0} = G|_{\mu, \Delta, \theta=0}$ is added for
regularization.
%In deriving
%the expression (\ref{bk3}), no assumptions
%was made concerning the
%smoothness of the phase variations, i.e., the expression is formally exact.

Let us now assume this smoothness
implying that
phase gradients are small. Then
$\Omega(\mu, T, \Delta(x), \partial \theta(x))$
can be approximated as follows:
\begin{equation}
\Omega (\mu, \Delta(x), \partial \theta(x))  \simeq
\Omega _{\rm kin} (\mu, T, \Delta, \partial \theta(x)) +
\Omega _{\rm pot} (\mu, T, \Delta),
\label{bk4}
\end{equation}
with the ``kinetic" term (see \cite{shar}, \cite{T})
\begin{equation}
\Omega _{\rm kin} (\mu, T, \Delta, \partial \theta(x))
 =  T \mbox{Tr} \sum_{n=1}^{\infty}
\left. \frac{1}{n} ({\cal G} \Sigma)^{n} \right|_{\Delta = \rm const}
\label{bkt5}
\end{equation}
and
the ``potential" term
\begin{equation}
\Omega _{\rm pot} (\mu, T, \Delta)  =
\left. \left(\frac{1}{g} \int d^D x \Delta^{2} -
T \mbox{Tr log} {\cal G}^{-1} +
T \mbox{Tr log} G_{0}^{-1} \right) \right|_{\Delta = \rm const}.
\label{bkt5a}
\end{equation}
The latter coincides with
 our earlier
mean-field energy (see also Appendix A),
 determining
the modulus of $\Delta (\mu , T)$
and thus
the stiffness of phase fluctuations.
The kinetic part  $\Omega_{\rm kin}$ contains gradient
terms whose size is determined by
the modulus of $\Delta (\mu , T)$.
Given the stiffness,
one may immediately set up an equivalent
XY-model.
 Both $\Omega_{\rm kin}$ and $\Omega_{\rm pot}$
are expressed in terms of the Green function of the fermions,
which solves the equation
%\mn{you correctly noted lack of that term in this equation
%%this is due to we  our approximation where
%modulus is independent of phase fluctuation }
\begin{equation}
\left[-\hat I \partial_{\tau} +
\tau_{3} \left(\frac{\nabla^{2}}{2 m} + \mu \right)
+ \tau_{1} \Delta \right]
{\cal G}(\tau, \mbox{\bf x}) = \delta(\tau) \delta(\mbox{\bf x})
\label{bkt6}
\end{equation}
and
\begin{equation}
\Sigma(\partial \theta) \equiv
\tau_{3} \left[ \frac {i \partial_{\tau} \theta}{2}  +
\frac{(\nabla \theta)^{2}}{8 m} \right] -
\hat{I} \left[\frac{i \nabla^{2} \theta}{4 m} +
\frac{i \nabla \theta(\tau, \mbox{\bf x}) \nabla}{2m} \right].
\label{bkt7}
\end{equation}
The gradient expansion that we use to determine stiffness
 was first made in Ref. ~\cite{shak} at zero temperature.
In Ref.~\cite{shar}, the kinetic term $\Omega_{\rm kin}$ was
calculated at finite temperature for arbitrary chemical
potential  retaining terms with $n= 1,2$
in the expansion (\ref{bkt5}).

 The result is (see Appendix A)
\begin{equation}
\Omega _{\rm kin} =
\frac{T}{2}
\int_{0}^{\beta} d \tau \int d^D x
\left[ n(\mu, T, \Delta ) i \partial_\tau \theta +
J(\mu, T, \Delta(\mu, T)) (\nabla \theta)^{2} +
K(\mu, T, \Delta(\mu, T)) (\partial_{\tau} \theta)^{2}
\right],
\label{bk8}
\end{equation}
where $J(\mu, T, \Delta)$ is the stiffness coefficient
\begin{equation}
J(\mu, T, \Delta) = \frac{1}{4m} n(\mu, T, \Delta) -
\frac{T}{4\pi} \int_{-\mu/2T}^{\infty} dx
\frac{x + \mu/2T}{\cosh^{2} \sqrt{x^{2} +
{ \Delta^{2}}/{ 4 T^{2}}}},
\label{bk9}
\end{equation}
\begin{equation}
K(\mu, T, \Delta) = \frac{m}{8 \pi} \left(
1 + \frac{\mu}{\sqrt{\mu^2 + \Delta^2}}
\tanh{\frac{\sqrt{\mu^2 + \Delta^2}}{2T}} \right),
\label{bk10}
\end{equation}
and   $n(\mu, T, \Delta)$
is the  density of fermions (\ref{numb})
which varies with temperature in our model.
At the temperature $T^*$ where
the modulus of $\Delta$ vanishes, also the stiffness
disappears.
The kinetic term
corresponds to an XY-model
with a Hamiltonian
\cite{bkt}, \cite{GFCM}:
\begin{eqnarray}
H={J \over 2} \int d {\bf x} [\nabla \theta({ \bf x})]^2,
\end{eqnarray}
the only difference with the
standard XY-model
lying in the
dependence of the stiffness constant $J$ on the
temperature, which is
determined from the solutions of
gap and number
equations (\ref{1.1}) and (\ref{1.2}).
Clearly, in this model the  Kosterlitz-Thouless transition always take
place below $T^*$.
In the XY-model with vortices of a high fugacity,
the temperature
of the phase transition is determined by
a simple formula \cite{Min}:
\begin{eqnarray}
T_{\rm KT}={\pi \over 2}J  \label{KT}
\end{eqnarray}
which follows from the
divergence of the average square size
of a vortex-antivortex pair.
Since these attract each other by a Coulomb potential
$v(r)=2\pi J\log(r/r_0)$,
the average square distance is
\begin{equation}
<r^2>\propto \int_{r_0}^\infty dr r \,r^2 e^{-(2\pi J/T)\log (r/r_0)}
\propto \frac{1}{4-2\pi J/ T},
\label{@}\end{equation}
which diverges indeed at the temperature (\ref{KT}).
In our case $T_{KT}$ should be determined self-consistently:
\begin{equation}
T_{\rm KT}=\frac{\pi}{2} J(\mu, T_{\rm KT}, \Delta(\mu, T_{\rm KT})).
\label{e1}
\end{equation}
{}From (\ref{bk9}),  (\ref{numb}) and     (\ref{KT})
it is easily seen that $T_{\rm KT}$ indeed tends to zero when
the pair attraction vanishes in which case $\Delta (T=0) = 0$.
In general, the behavior
 of $T_{\rm KT}$
for strong and weak couplings
is found by the following considerations.
We observe
from the above-derived
limiting formulas
for $\Delta(T,\mu)$ and $T^*$,
that the particle number $n$
does not vary appreciably in these limits
with temperature in the range $0 < T < T^*$,
so that weak-coupling
estimates for $T_{\rm KT}$ derived
within the model with temperature-independent chemical potential
(i.e. when the system is coupled to a large reservoir,
see discussion after the formula (\ref{1.14}) ) practically
coincide with those derived from a
fixed fermion density.
Further it is
immediately realized from the equations
(\ref{e1}),
(\ref{numb}),
(\ref{bk9}) and
(\ref{1.3.2})
that
in the weak-coupling limit
$\Delta (T_{\rm KT},\mu)/T_{\rm KT}$ is a small
parameter.
At zero coupling, the stiffness $J(\mu,T_{\rm KT},\Delta (\mu, T_{\rm KT}))$
vanishes identically, such that  an estimate
of $J$ at  weak couplings
requires
calculating a lowest-order
correction
to the second term of eq.(\ref{bk9})
 proportional to
 $\Delta(T_{\rm KT}, \mu)/T_{\rm KT} $.
Thus weak-coupling expression for
stiffness reads \cite{shar}:
\begin{equation}
J(T) \simeq \frac{7 \zeta (3)}{ 16 \pi^3} \epsilon_F
\frac{\Delta(T)^2}{T^*{}^2}.
\end{equation}
Using the limiting behavior  (\ref{1tc1})
of $\Delta(T)$
[see also discussion  after
 (\ref{tcstr3d})]
we find after some algebra
 the weak-coupling equation for $T_{\rm KT}$:
\begin{equation}
T_{\rm KT} \simeq
\frac{ \epsilon_F}{4}
\left( 1- \frac{T_{\rm KT}}{T^*} \right).
\label{e2}
\end{equation}
where  $\epsilon_F =(\pi / m) n$
is
the
Fermi energy of free fermions.
It is useful to introduce reduced dimensionless temperatures
${\tilde T_{KT}} \equiv T_{KT} / \epsilon_F$ and
${\tilde T^*} = T^* / \epsilon_F$
which are small in the weak-coupling limit.
Then we rewrite Eq.~(\ref{e2}) as
\begin{equation}
{\tilde T_{KT}} \simeq
\frac{1}{4}
\frac{1}{1+1 / 4{\tilde T^*}}.
\end{equation}
For small ${ \tilde T}^*$ we may expand
\begin{equation}
{\tilde T_{KT}} \approx {\tilde T^*} - \frac{ {\tilde T^*}{}^2}{4}.
\label{s01}
\end{equation}
This equation shows
 nicely how for
decreasing coupling strength
$T_{\rm KT}$
merges with $T^*$.

\old{In  three dimensions,
a qualitatively similar
behavior is known from BCS theory where
these temperatures coincide.
In the papers
 \cite{Noz}, \cite{R8},
these merging in three dimensions was obtained
by considering
the condensation of a
boson gas formed from the fermion pairs
retaining corrections to the number equation.
We will discuss it in the next section}

\old{to the ordinary BCS superconductivity
where thermal destruction of pairs determines $T_c$.
For strong couplings in 3D onset of superconductivity
is controlled by thermal excitation of the
collective modes that destruct phase coherence
well below temperature of the thermal destruction of pairs.}
\old{
 Thus we have shown that
in the weak-coupling limit of the two dimensional
transition, the temperature  $T_{\rm KT}$
of quasicondensate formation tends
to the temperature of $T^*$ of pair
formation, and can thus be interpreted
like in three-dimensional case \cite{Noz,R8}
that in the weak-couplings limit
in two-dimensional system critical
temperature is determined by
thermal destruction of composite
bosons and thus can be described well in this
limit by ordinary two-dimensional BCS mean-field equations.
Apparently, a pseudogap phase ceases to exist in this limit.
}

As a function of the
crossover parameter $x_0$,
the temperature
$T_{\rm KT}$
behaves like
\begin{eqnarray}
{\tilde T}_{\rm KT} \approx
\frac{e^\gamma}{\pi} \frac{1}{x_0}.
\label{c1}
\end{eqnarray}
The merging of the two temperatures
in the  weak-coupling regime is displayed
in Fig.~\ref{f7}.
\begin{figure}[tbh]
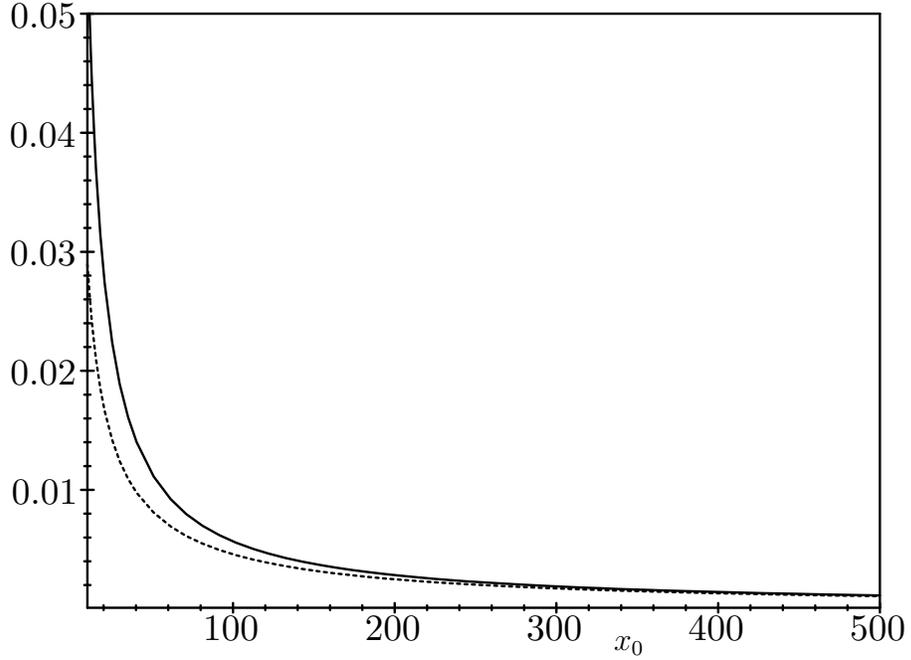
%##
\input kt.tps  \\[-2cm]
\caption{
Weak-coupling behavior of $T_{KT}(x_0)$.
The solid line is $T^*$, the
dashed line represent $T_{KT}$.
}
\label{f7}\end{figure}

Consider now the opposite limit of strong couplings.
There Eqs. (\ref{e1}),
(\ref{numb}),
(\ref{bk9}), and
(\ref{1.3.2})
for $T_{\rm KT}$, $n(T, \mu)$, and $\Delta (T, \mu)$ show that
 $T_{\rm KT}$ tends to a constant value.
{}From Eqs. (\ref{bk9}),
(\ref{1.3.2}), (\ref{numb}), and
(\ref{e1}), as well as the limiting expressions
 (\ref{2dbe})
and (\ref{extr}) it follows that in the
strong-coupling limit $\Delta(T_{\rm KT})$ is always
situated close to
the zero-temperature value of
$\Delta (T_{\rm KT}, \mu) \approx \Delta (T=0,\mu )$.
Taking this into the account we derive
an estimate for the second term in (\ref{bk9}),
thus obtaining the strong-coupling equation for $T_{\rm KT}$:
\begin{eqnarray}
T_{\rm KT}
\simeq \frac{\pi}{8} \left\{
\frac{1}{m}n -
\frac{T_{\rm KT}}{\pi} \exp\left[
 \frac{\sqrt{\mu^2+\Delta^2(T_{\rm KT},\mu)}}{T_{\rm KT}}
\right]
\right\}
%\nonumber
\label{e30}
\end{eqnarray}
which may further be expanded
as
\begin{eqnarray}
T_{\rm KT}
\simeq
 \frac{\pi}{8} \left\{
\frac{1}{m}n - \frac{T_{\rm KT}}{\pi} \exp\left[  \frac{\mu}{T_{\rm
KT}}
+ \frac{\Delta^2 (T_{\rm KT}, \mu)}{2 \mu T_{\rm KT}}  \right]
\right\}.
\label{e3}
\end{eqnarray}
With the
approximation
$\Delta (T_{\rm KT}, \mu) \approx \Delta (T= 0, \mu)$
and the limiting behavior (\ref{1.13})
we find that the first term in the exponent tends
in the strong-coupling limit to a constant,
$ \Delta^2 (T_{\rm KT}, \mu)/2 \mu T_{\rm KT} \rightarrow - 4$,
whereas the first  term in the brackets
tends to $- \infty$,
so that Eq.~(\ref{e3})
has the limiting form
\begin{eqnarray}
T_{\rm KT} \simeq
\frac{\pi}{8} \frac{n}{m}
\left\{
1 - \frac{1}{8} \exp\left[
\frac{2\mu}{\epsilon_F} -4
\right]
\right\}.
\label{e32}
\end{eqnarray}
As a function of $x_0$ crossover parameter, this
reads
\begin{eqnarray}
T_{\rm KT} \simeq
\frac{\pi}{8} \frac{n}{m}
\left\{
1 - \frac{1}{8} \exp\left[
8x_0^2 -6
\right]
\right\}.
\label{e32a}
\end{eqnarray}
 Thus  for increasing coupling strength, i.e., decreasing
crossover parameter $x_0 \ll -1$,
the phase-decoherence temperature $T_{\rm KT}$
tends very quickly towards a constant:
\begin{equation}
T_{KT} \simeq  \frac{\pi}{8} \frac{n}{m}.
\label{e302}
\end{equation}
In this limit
we know from Eq.~(\ref{numb}) that the difference
in the carrier density at zero temperature, $n(T=0)$,
becomes equal to $n(T=T_{KT})$,
so that our
limiting result coincides
with that obtained in
the "fixed carrier density model":
\begin{equation}
T_{\rm KT} = \frac{\epsilon_F(n_0)}{8} = \frac{\pi}{8m} n_0,
\label{e4}
\end{equation}
where we have inserted again $\epsilon_F(n)=(\pi/m)n$
for the
Fermi energy of free fermions
at the carrier density
$n_0=n(T=0)$.

Note that
this strong-coupling
behavior
of $T_{\rm KT}$
coincides
roughly
with the estimate
of $T_c$ for
three-dimensional
superconductors in Refs.~ \cite{Noz}, \cite{R8}, and \cite{h}
via
the
onset of Bose condensation
of tightly bound, almost free composite bosons.
In the first two of these references
which include only quadratic fluctuations around the mean field
(corresponding to ladder diagrams),
$T_c$ was shown to tend
to a constant value
% $T_c=[n/2 \zeta(3/2)]^{2/3} \pi /m$,
which does not depend on the internal
structure of composite boson and
 is simply equal
to the
condensation
temperature of
a gas of free bosons
of mass $2m$ and density
$n/2$, implying that the interactions
between the
composite bosons is irrelevant in this
approximation.

We find the same situation in two dimensions,
where $T_{KT}$ tends to a constant depending only on
the mass $2m$ and the density $n/2$ of the pairs.
No dependence on the coupling strength is left.
The difference with respect to the three-dimensional case
is that here the
transition temperature $T_c=T_{KT}$
is
linear in the carrier density $n$, while growing like $n^{2/3}$
in three dimensions.
Our result (\ref{e4}) agrees
with
Ref.~\cite{Dr} and ~\cite{shar}.%

\old{\footnote{ Surprisingly, a linear dependence of the
critical temperature on $n$ is
a behavior which has been observed experimentally
in cuprates even though there are a number of evidences
that optimally doped  and underdoped cuprates
are still on the weak-coupling side of crossover.}}

If interactions
between condensed and noncondensed composite
are taken into account,
as done in Ref.~\cite{h}
 in three dimensions, then
 $T_c$
turns out to grow slowly with the coupling
strength in the strong coupling regime.

Equation (\ref{e4})
determines
the
critical temperature in the strong-coupling
limit completely in terms of
the carrier density $n_0$.
There exists a
corresponding equation
for the temperature $T^*$
in the strong-coupling
limit
$\epsilon_0 \gg \epsilon_F$:
\begin{equation}
T^* \simeq \frac{\epsilon_0}{2}\frac{1}{ \log\lfrac{\epsilon_0}{\epsilon_F}}.
\end{equation}

For experimental purposes,
the dependence
of the ratio
$2 \Delta(0) / T_{KT}$
on the coupling strength
is of interest. It is plotted
  in Fig.~\ref{f8}.
Analytically, we have
in the weak-coupling
limit
\begin{equation}
\frac{2 \Delta(0)}{T_{KT}} =
\frac{2 \pi}{e^\gamma}
\left\{
1+\frac{e^\gamma}{\pi}\frac{4}{x_0} +
\left[
\frac{1}{8} +
\left(
\frac{4 e^{\gamma}}{\pi}
\right)^2
\right]\frac{1}{x_0^2}
\right\} +{\cal O}\left(x_0^{-3}\right)  ,
\end{equation}
and for strong-couplings:
\begin{equation}
\frac{2\Delta(0)}{T_{KT}} \simeq
\frac{32}{\sqrt{x_0^2+1}+x_0} \simeq -64 x_0
{}.
\end{equation}
The two curves can easily be interpolated graphically
to all
coupling strengths, as seen in the figure.
\begin{figure}[tbh]
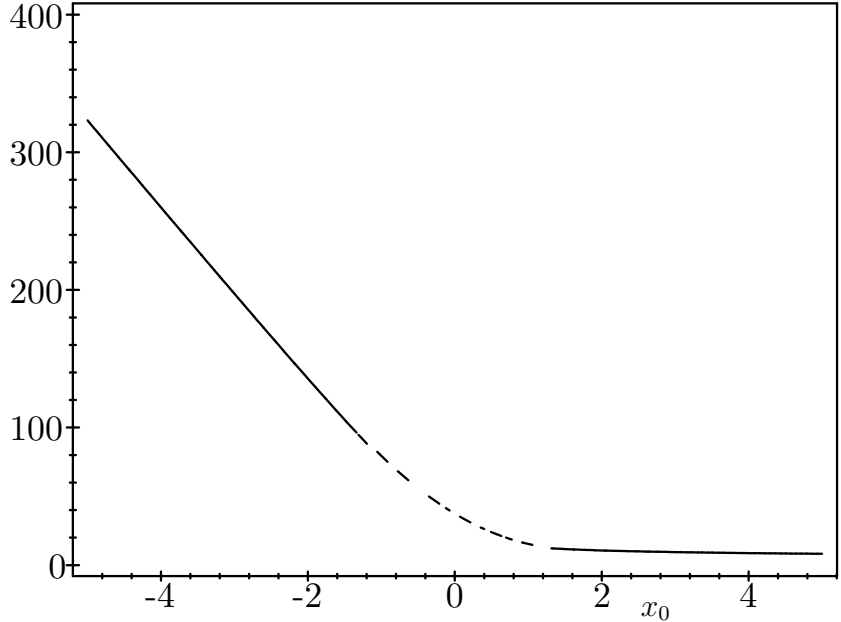
%##
\input 8.tps  \\[-2cm]
\caption{
Weak-coupling  and strong-coupling
estimates for the ratio
 $2\Delta (0)/T_{KT}$ (solid curves).
The dashed line is a graphical interpolation.
}
\label{f8}\end{figure}

\section[Phase Fluctuations in Three Dimensions and Superconductuctive
Transition in 3D XY-Model]
{Phase Fluctuations in Three Dimensions and Super\-conductuctive Transition in
3D XY-Model}

 In this chapter we discuss effects of
fluctuations in three-dimensional systems.
In three dimensions we have the expression for stiffness
\begin{equation}
J_{\rm 3D}(\mu,T, \Delta)
=\frac{1}{4m}n(\mu,T,\Delta)-
\frac{ \sqrt{2m}}{16 \pi^2}\frac{1}{T}
\int^\infty_{-\mu}
d \xi
\frac{(\xi+\mu)^{3/2}}{\cosh^2(\sqrt{\xi^2+\Delta^2}/2T)}
\end{equation}
With it we can immediately set up 3D XY model,
whose Hamiltonian reads:
\begin{equation}
H= \frac{ J_{\rm 3D}}{2} \int d^3 { \bf x} [ \nabla \theta( {\bf x})]^2.
\end{equation}
Temperature of the phase transition of this model can
be estimated using mean-field methods for the lattice 3D XY-Model
\cite{GFCM}:

\begin{equation}
T_{\rm 3D}^{MF} \simeq 3 J_{\rm 3D}a,
\end{equation}
 $a=1/n_b^{1/3}$ is the lattice spacing of the theory \cite{GFCM}
where $n_b$ is number of pairs.

In the weak-coupling limit, the stiffness can be expanded
near $T^*$ as follows:
\begin{equation}
J=\frac{7}{32 \pi^4} \zeta(3) \frac{p_F^3}{m} \frac{\Delta^2}{T^*{}^2},
\end{equation}
This  is similar
to  the coefficient
of the gradient term in the Ginzburg-Landau expansion
except that there
it is obtained from
 a
small-$\Delta$   expansion,
whereas here the background gap has a nonzero modulus.

Obviously we have no separation of the two temperatures in the BCS limit.
For moderately strong coupling,
the two temperatures are related by
\begin{equation}
{\tilde T_c }= {\tilde T^* } - \alpha {\tilde T^*{}^{5/2}},
\end{equation}
where $\alpha=(2 \pi^2)^{2/3}/3$.

In the strong-coupling limit of the theory
where we have tightly bound composite bosons,
the phase stiffness tends asymptotically to:
\begin{equation}
J=\frac{n}{4m} - \frac{3 \sqrt{2 \pi m}}{16 \pi^2} T^{3/2}
\exp\left[-\frac{\sqrt{\mu^2+\Delta^2}}{T}
\right],
\label{@stiff@}
\end{equation}
which may be expanded
\begin{equation}
J=\frac{n}{4m} - \frac{3 \sqrt{2 \pi m}}{16 \pi^2} T^{3/2}
\exp\left[\frac{\mu}{T}
\right].
\label{@stiff2@}
\end{equation}
It obviously tends in this limit quickly to
\begin{equation}
J=\frac{n}{4m}.
\end{equation}
An estimate for the critical temperature, obtained
via the
mean-field treatment of the 3D XY-model
on the
lattice reads in this limit:
\begin{equation}
T_c= 
\frac{3}{2m} \left[ \left(\frac{n}{2}\right)^{2/3}-
\frac{1}{n^{1/3} } \frac{2^{7/6}}{\pi^{3/2}} T_c^{3/2} m^{3/2}
\exp\left( -\frac{\sqrt{\mu^2+\Delta^2}}{T_c}
\right)
\right]
%\nonumber
%\\
%\simeq  \frac{3 n^{2/3}}{2^{5/3}m}-
%\frac{3}{2^{1/6} 8 \pi^{3/2} } \frac{1}{\sqrt{m}n^{1/3}}
%\exp\left( \frac{\mu}{T}
%\right).
\label{chaoslabs}
\end{equation}
This quickly tends from below to the  value:
\begin{equation}
T_c^{ \rm 3D XY} = \frac{3 n^{2/3}}{2^{5/3} m}=
\epsilon_F \frac{3}{(6 \pi^2)^{2/3}} \simeq 0.198 \epsilon_F.
\end{equation}
This result is very close to
the temperature of the condensation of bosons
of mass $2m$
and density $n/2$, which, as it was discussed in the introduction
was obtained including
the effect of Gaussian fluctuations
into the mean-field equation for the particle number \cite{Noz,R8}
yielding \footnote{When critical temperature
is studied via retaining gaussian corrections to the number
equation the crossover of the critical temperature
has an artificial maximum in the region of intermediate couplings
\cite{Noz,R8}, so in this
case it's limiting value is approached from above in the strong-coupling limit,
 this is not the case in our approach. }
\begin{equation}
T_c^{\rm Bosons} =[n/2\zeta(3/2)]^{2/3} \pi/m =0.218 \epsilon_F.
\label{@bosons}
\end{equation}

\old{There is another way to deal with
superconductive transition of 3D XY model,temperature
of the
bose condensation of composite bosons can be found with help of
so called vortex proliferation transition on the lattice.
The
Kosterlitz-Thouless transition in two dimensions
has a counterpart in three dimensions in a vortex proliferation transition.
At this point, the
configurational entropy of
vortex lines overcomes the Boltzmann suppression of these vortex line.
There exists an analogous dual mechanism
if the transition is approached from above the
critical temperature.
There
the
normal
state performs
fluctuations into the superconducting state
by
occasionally creating small rings of superflow,
which are closed due to current conservation.
Close to critical point these rings grow in size,
becoming
infinitely long
at the transition point,
 again due to their
configurational entropy.}

The XY-model nature of the phase transition at $T_c$
has been demonstrated
in recent
experiments
\cite{3dxy}  on YBa$_2$Cu$_3$O$_{7-\delta}$
near the region of optimal doping.
\old{Let us therefore briefly recall where
the transition takes place in the
three-dimensional XY-model.
 For proliferation transition
its Hamiltonian has the form
\begin{equation}
H= \frac{J_{\rm prol}}{2} \int d^3 { \bf x} [ \nabla \theta( {\bf x})]^2,
\end{equation}
where $J_{\rm prol}$ is the stiffness
constant of phase fluctuations.}
The phase transition of this model was discussed in great detail on a
lattice in
the textbook \cite{GFCM}, since it describes
the critical properties of
the superfluid transition of Helium.

\section{ Conclusion}
We have discussed the two different
transitions taking place in superconductors at
strong couplings,
the formation of pairs
and onset of phase coherence.
For theory it was a fortunate fact of history
that the early-discovered
metallic superconductors
had such a weak
 coupling that there was only one transition
which, moreover, can be understood by mean-field methods.
In
high-$T_c$
superconductors, the existence of a pseudogap
brings in complications
which we have tried to illuminate
in the framework of the simple fermion model with
$ \delta $-function interaction.
The interpretation of the experimental data
is still complicated due to
the complex chemical structure of these compounds.
Little is known
up to now
on the real
forces causing the
pairing.

We have studied
the crossover
from BCS-type to Bose-type
superconductivity and the behavior of the pseudogap state.
Our crossover parameter is
 $x_0$, a quantity
 closely related to the chemical potential.
For this purpose
we have used the gradient expansion
of the effective energy functional
to set up an equivalent XY-model
which allows us to investigate
the onset of long-range order in the phase fluctuations.
In two dimensions, we have given
a simple analytic expression
which shows
how
the resulting
Kosterlitz-Thouless temperature
$T_{\rm KT}$
at which quasi-long-range order sets in
 moves towards the pair-binding temperature
$T^*$, and merges with it in the weak-coupling limit.
\old{In this limit,
the temperature of onset of
the superconductive phase
coincides with
the onset of pair formation,
and
the pseudogap phase
disappears---the familiar scenario of
 of ordinary  BCS
superconductivity.}
\comment{This should
be expected
on the basis of the
 equivalence
between fermions and vortices in low-dimensional field theories,
which make
the thermal breaking
of vortex pairs
  a bosonic description of the
breakup of electron pairs (the antivortex in each pair
describing an electron below the Fermi surface).}
We have found similar results
in three dimensions, setting up a
three dimensional
$XY$-model for the description of the
onset of phase coherence in the
superconductive transition,
and how this transition
evolves
to the ordinary BCS transition
in the weak-coupling regime.

We have also  studied the weak- to strong-coupling crossover
of thermodynamic functions
of the superconductive state near zero temperature as well
in the pseudogap phase
near the critical temperature.
\old{Even though
an oversimplified model of electrons with $ \delta$-function attraction
was used which cannot account for the
individual microscopic
properties of realistic solid-state systems,
the
qualitative behavior
of all calculated properties
in the weak-coupling
regime should have relevance for understanding behavior of some
high-$T_c$
superconductors.
For strong-couplings, the
pseudogap behavior is of independent interest.}
\old{
In contrast to BCS superconductivity,
where Cooper pairs are separated at
the critical temperature $T_c$,
the critical temperature in the
strong-coupling regime
is characterized by the onset
of long-range
phase coherence in an ensemble
of Bose-Einstein particles
formed by bound electron pairs.}

Certainly, our
mean-field estimates
for $T^*$
are quite crude, and we expect
significant modifications
due to
fluctuations, in particular of the
character of the transition
which experimentally does not seem to be of
second order, and may not be a phase transition
after all.
All our formulas for thermodynamic quantities
will have to be smeared out
in temperature near $T^*$,
before any possible  comparison with experiments.

\old{
In the strong-couplings limit
we suggest also in three dimensions
the relevance of the
XY-model for estimating the critical temperature $T_c$,
which we identify with the
vortex proliferation temperature $T_{ \rm prol}$
in superfluids.}
\old{
If the repulsion between the strongly bound fermion pairs
were weak, the temperature $T_{\rm prol}$
would coincide with the well-known transition temperature of a
Bose gas.}

Let us finally remark that the separation of $T^*$
and $T_c$ has an  analogy
in the ferroelectrics and magnets
which also contain two
separate characteristic temperatures, for example in the later case---
the Stoner- and the Curie-temperature.

\section{Acknowledgments}
We thank Profs. K. Bennemann, K. Maki, and V.Emery  for
explaining to us some aspects of $T^*$ crossover in superconductive cuprates.
One of us (E.B.) is grateful to all members of Prof.~Kleinert's
group at the  Institut f\"ur Theoretische Physik
Freie Universit\"at Berlin for their kind
hospitality, and
to Drs. A.V. Goltsev, S.A. Ktitorov, B.N. Shalaev, S.G. Sharapov
and Profs. Yu.A. Firsov and V.M. Loktev
for discussions of our results.

\appendix
\section{Action functional of Collective Pair Field\label{App.A}}
In this appendix we briefly outline derivation
of the effective action (\ref{dfg}).
As shown in Ref.~\cite{6'}, a pair field $\bf \Delta$ is
introduced to eliminate the quartic interaction term in the
 functional integral
involving the
action of the Hamiltonian (\ref{1.0}):
\begin{eqnarray}
{\cal A}= \int dt\left[  \sum_\sigma \int \! d^D x
        \, \psi_\sigma^{\dag} ({\bf x})i\hbar \partial _t
        \psi_\sigma({\bf x})
         -               H(t)\right]
        \label{1.0b},
\end{eqnarray}
After that, the fermions can be integrated out. At a constant
pair field, we find the potential part
(\ref{bkt5a}) of the
collective-field action
\begin{eqnarray}
&&\!\!\!\!\!\!\!\!\!\!\!\!\!\!\!\!\!\!\!\!\!\!
\!\!\!\!\!\!\!\!\!\!\!\!
 \Omega_{\rm pot}( \mu, T, {\bf \Delta},{\bf \Delta}^{\ast}) = V \left\{
\frac{|{\bf \Delta}|^{2}}{g} - T \sum_{n = -\infty}^{+\infty}
\int \frac{d^D k}{(2 \pi)^D}
\mbox{tr} [\ln G^{-1} (i \omega_{n}, \mbox{\bf k})
e^{i \delta \omega_{n} \tau_{3}}]
\right.
\nonumber                                \\
&& \qquad
+ \left. T \sum_{n = -\infty}^{+\infty}
\int \frac{d^D k}{(2 \pi)^D}
\mbox{tr} [\log G_{0}^{-1} (i \omega_{n}, \mbox{\bf k})
e^{i \delta \omega_{n} \tau_{3}}] \right\}, \quad
\delta \to +0,                                 \label{ab1}
\end{eqnarray}
where ${\bf \Delta}= \Delta e^{i \theta}$ and
\begin{equation}
G^{-1} (i \omega_{n}, \mbox{\bf k}) =
i \omega_{n} \hat I - \tau_{3} \xi(\mbox{\bf k}) +
\tau_{+} {\bf \Delta} + \tau_{-} {\bf \Delta}^{\ast} =
\left( \begin{array}{ccc}
i \omega_{n} - \xi(\mbox{\bf k}) & {\bf \Delta} \\
{\bf \Delta}^{\ast}                      & i \omega_{n} + \xi(\mbox{\bf k})
\end{array} \right)                          .   \label{ab2}
\end{equation}
%and
%\begin{equation}
%G_{0}^{-1} (i \omega_{n}, \mbox{\bf k}) =
%\left. G^{-1} (i \omega_{n}, \mbox{\bf k})
%\right|_{{\bf \Delta} = {\bf \Delta}^{\ast} = \mu =0} =
%\left( \begin{array}{ccc}
%i \omega_{n} - \varepsilon(\mbox{\bf k}) & 0 \\
%0                      & i \omega_{n} + \varepsilon(\mbox{\bf k})
%\end{array} \right)                             \label{ab3}
%\end{equation}
Using the
 identity $\mbox{tr} \log \hat A = \log \det \hat A$,
equation
(\ref{ab1}) takes the form
\old{\begin{eqnarray}
\Omega_{\rm pot}( \mu, T, {\bf \Delta},{\bf \Delta}^{\ast}) = V \left\{
\frac{|{\bf \Delta}|^{2}}{g} \right. & - & T \sum_{n = -\infty}^{+\infty}
\int \frac{d^D k}{(2 \pi)^D}
\log \frac{\det G^{-1} (i \omega_{n}, \mbox{\bf k})}
{\det G_{0}^{-1} (i \omega_{n}, \mbox{\bf k})}
\nonumber                                \\
& - & \left. \int \frac{d^D k}{(2 \pi)^D}
[-\xi (\mbox{\bf k}) + \varepsilon (\mbox{\bf k})] \right\} =
                                          \label{ab4}
\end{eqnarray}}
\begin{eqnarray}
\Omega_{\rm pot}( \mu, T, {\bf \Delta},{\bf \Delta}^{\ast}) = V \left\{
\frac{|{\bf \Delta}|^{2}}{g} \right. & - & T \sum_{n = -\infty}^{+\infty}
\int \frac{d^D k}{(2 \pi)^D}
\log \frac{\omega_{n}^{2} + \xi^{2}(\mbox{\bf k}) + |{\bf \Delta}|^{2}}
{\omega_{n}^{2} + \varepsilon^{2}(\mbox{\bf k})}
\nonumber                                \\
& - & \left. \int \frac{d^D k}{(2 \pi)^D}
[-\xi (\mbox{\bf k}) + \varepsilon (\mbox{\bf k})] \right\},
                                          \label{ab5}
\end{eqnarray}
After performing the sum over the Matsubara frequencies
in (\ref{ab5}),
we obtain the well-known mean-field expression
for $\Omega_{\rm pot}$ \cite{6'}:
\begin{eqnarray}
&&\Omega_{\rm pot}(\mu, T, {\bf \Delta}, {\bf \Delta}^{\ast})  =
V \left\{\frac{|{\bf \Delta}|^{2}}{g} \right.  -
 \int \frac{d^D k}{(2 \pi)^D}
\left[ 2 T
\log 2 \cosh \frac{\sqrt{\xi^{2}(\mbox{\bf k}) + |{\bf \Delta}|^{2}}}{2T}
- \xi(\mbox{\bf k}) \right]
\nonumber                        \\
&& \qquad \qquad            ~~~~~~~~~~~+
\left.   \int \frac{d^D k}{(2 \pi)^D}
\left[ 2 T
\log 2 \cosh \frac{\varepsilon(\mbox{\bf k})}{2T}
- \varepsilon(\mbox{\bf k}) \right] \right\}.
\label{ab6}
\end{eqnarray}
In order to derive the kinetic part
 $\Omega_{\rm kin}$ of the mean-field energy,
we must calculate the first two terms
of the series (\ref{bkt5}), the first being \cite{shar}
\begin{equation}
\Omega_{\rm kin}^{(1)} = T \int_{0}^{\beta} d \tau \int d^D x
\frac{T}{(2 \pi)^{2}} \sum_{n = - \infty}^{\infty}
\int d^D k
\,\mbox{tr} [{\cal G}(i \omega_{n}, \mbox{\bf k}) \tau_{3}]
\left[ \frac{i \partial_{\tau} \theta}{2} +
\frac{(\nabla \theta)^{2}}{8 m}\right],         \label{b1bk1}
\end{equation}
with
\begin{equation}
{\cal G}(i \omega_{n}, \mbox{\bf k}) = - \frac{
i \omega_{n} \hat{I} + \tau_{3} \xi(\mbox{\bf k}) - \tau_{1} \Delta}
{\omega_{n}^{2} + \xi^{2}(\mbox{\bf k}) + \Delta^{2}}     .\label{b1bk2}
\end{equation}
After summing over the Matsubara
frequencies and integration over
$\mbox{\bf k}$, we obtain
\begin{equation}
\Omega_{\rm kin}^{(1)} =
T \int_{0}^{\beta} d \tau \int d^D x
\,n(\mu, T, \Delta) \,\left[
\frac{i \partial_{\tau} \theta}{2} + \frac{(\nabla \theta)^{2}}{8 m}
\right],                \label{b1bk3}
\end{equation} with $n(\mu, T, \Delta)$ given in by (\ref{1.2}).
After an ansatz
$\Sigma = \tau_{3} O_{1} + \hat{I} O_{2}$,
where $O_{1}$ and
$O_{2}$ are the two gradient terms in Eq.~(\ref{bkt7}),
we derive for the second term
$\Omega^{(2)}_{\rm kin}$
the two contributions from
$O_{1}$ and
$O_{2}$:
\begin{equation}
\Omega_{\rm kin}^{(2)} (O_{1}) = - \frac{T}{2}
\int_{0}^{\beta} d \tau \int d^D x
\,K(\mu, T, \Delta)
\left[ \frac{i \partial_{\tau} \theta}{2} +
\frac{(\nabla \theta)^{2}}{8 m}\right]^{2},         \label{b1bk4}
\end{equation}
where $K(\mu, T, \Delta)$ was given in (\ref{bk10}).
This is the second term in (\ref{bk8}).
The second term in (\ref{bk9})
is obtained from the second contribution to $\Omega_{\rm kin}^{(2)} $:
\begin{equation}
\Omega_{\rm kin}^{(2)} (O_{2}) = -
\int_{0}^{\beta} d \tau \int d^D x
\frac{1}{32 \pi^{2} m^{2}}
\int d^D k
\frac{\mbox{\bf k}^{2}}
{\cosh^{2} [{ \sqrt{\xi^{2}(\mbox{\bf k}) + \Delta^{2}}}/{2T}]}
(\nabla \theta)^{2}.                                    \label{b1bk5}
\end{equation}

 Combining (\ref{b1bk5}), (\ref{b1bk4}) and (\ref{b1bk3})
we obtain (\ref{bk8}).

\vskip 0.6cm
%\begin{center}
%\end{center}
\vskip 0.4cm
%\newpage

\end{document}